\documentclass{tcibook}
\usepackage{fancyhea}
\usepackage{work}
\usepackage{bm}       
\usepackage{graphicx}
\usepackage{hyperref}      
\usepackage{amssymb}

\newcommand{\nc}{\newcommand}  

\def\beq{\begin{equation}}
\def\eeq#1{\label{#1}\end{equation}}
\def\eeqn{\end{equation}}

\newenvironment{Eqnarray}%
   {\arraycolsep 0.14em\begin{eqnarray}}{\end{eqnarray}}
\def\beqa{\begin{Eqnarray}}
\def\eeqa#1{\label{#1}\end{Eqnarray}}
\def\eeqan{\end{Eqnarray}}

\nc{\ra}{\rightarrow}  
\nc{\slsh}{\slash\hspace*{-0.22cm}}
\def\Re{{\cal R \mskip-4mu \lower.1ex \hbox{\it e}\,}}
\def\Im{{\cal I \mskip-5mu \lower.1ex \hbox{\it m}\,}}

\nc{\vev}[1]{ \left\langle {#1} \right\rangle }
\nc{\bra}[1]{ \langle {#1} | }
\nc{\ket}[1]{ | {#1} \rangle }
\nc{\fb}{\,{\rm fb}^{-1}}
\nc{\ev}{{\rm eV}}
\nc{\kev}{{\rm keV}}
\nc{\Mev}{{\rm MeV}}
\nc{\gev}{{\rm GeV}}
\nc{\tev}{{\rm TeV}}
\nc{\mev}{{\rm MeV}}

\def\del{\partial}
\def\Dslash{\not{\hbox{\kern-4pt $D$}}}
\def\dslash{\not{\hbox{\kern-2pt $\del$}}}
\def\pslash{\not{\hbox{\kern-2pt $p$}}}
\def\ETmiss{ \not{\hbox{\kern-4pt $E$}}_T }

\def\msb{{\bar{\ssstyle M \kern -1pt S}}}

\newcommand{\mrm}[1]{\mathrm{#1}}
\newcommand{\ttt}[1]{\texttt{#1}}
\newcommand{\py}{\ttt{PYTHIA}}

\newcommand{\sib}{\ttt{SIBYLL}}
\newcommand{\qgs}{\ttt{QGSJET}}
\newcommand{\epos}{\ttt{EPOS}}

\newcommand{\figRef}[1]{fig.~\ref{#1}}

\newcommand{\ybco}{\ensuremath{\mathrm{YBa_2Cu_3O_{7-\delta}}}}

\begin{document}

\def\bibname{References}
\bibliographystyle{plain}

\raggedbottom

\pagenumbering{roman}

\parindent=0pt
\parskip=8pt
\setlength{\evensidemargin}{0pt}
\setlength{\oddsidemargin}{0pt}
\setlength{\marginparsep}{0.0in}
\setlength{\marginparwidth}{0.0in}
\marginparpush=0pt


\pagenumbering{arabic}

\renewcommand{\chapname}{chap:intro_}
\renewcommand{\chapterdir}{.}
\renewcommand{\arraystretch}{1.25}
\addtolength{\arraycolsep}{-3pt}

\def \lsim{\mathrel{\vcenter
     {\hbox{$<$}\nointerlineskip\hbox{$\sim$}}}}
\def \gsim{\mathrel{\vcenter
     {\hbox{$>$}\nointerlineskip\hbox{$\sim$}}}}

\chapter*{High Energy Hadron Colliders}
\label{chap:had}

\begin{center}\begin{boldmath}

\begin{center}


\begin{large} {\bf Frontier Capabilities - Hadron Collider Study Group
\footnote{This document was prepared as the final report of the Frontier Capabilities Hadron Collider 
Study Group at the 2013 Community Summer Study ``Snowmass 2013'', Minneapolis, MN, July-August 2013.}} \end{large}

W.~Barletta, 
M.~Battaglia, 
M.~Klute, 
M.~Mangano, 
S.~Prestemon, 
L.~Rossi,
P.~Skands

\end{center}

\end{boldmath}\end{center}

\section{Executive Summary}
\label{sec:intro}

High energy hadron colliders have been the tools for discovery at the highest mass scales of the 
energy frontier from the S$\bar{p}p$S, to the Tevatron and now the LHC.  
They will remain so, unchallenged for the foreseeable future. The discovery of the Higgs 
boson at the LHC, opens a new era for particle physics. 
After this discovery, understanding what is the origin of electro-weak symmetry 
breaking becomes the next key challenge for collider physics. This challenge can be expressed in terms 
of two questions: up to which level of precision does the Higgs boson behave like predicted by the SM? 
Where are the new particles that should solve the electro-weak (EW) naturalness problem and, possibly, offer 
some insight into the origin of dark matter, the matter-antimatter asymmetry, and neutrino masses? 
The approved CERN LHC programme, its future upgrade towards higher luminosities (HL-LHC), and the study of  
an LHC energy upgrade (HE-LHC) or of a new proton collider delivering collisions at a center of mass energy up to 
100~TeV (VHE-LHC), are all essential components of this endeavor.

{\sl The full exploitation of the LHC is the highest priority of the energy frontier, hadron collider program.} 

LHC is expected to restart in Spring 2015 at center-of-mass energy of 13-14 TeV and its design luminosity of 
10$^{34}$ cm$^{-2}$ s$^{-1}$ to be reached during 2015.  After 2020, some critical components of the accelerator will 
reach the radiation damage limit and others will have reduced reliability, also due to radiation effects. Furthermore, 
as the statistical gain in running the accelerator without substantially increased luminosity will become marginal, 
the LHC will need a decisive increase of its luminosity\cite{ATLAS-collaboration:2012iza}.  
This new phase of the LHC life, named the High Luminosity LHC (HL-LHC), will prepare the machine to attain the 
astonishing threshold of 3000 fb$^{-1}$ of integrated luminosity during its first decade of 
operation~\cite{Zimmermann:2011za}. High-luminosity offers the potential to increase the 
precision of several key LHC measurements, to uncover rare processes, and to guide and validate the progress 
in theoretical modeling, thus reducing the systematic uncertainties in the interpretation of the data.  
The project is now the first priority of Europe, as stated by the Strategy Update 
for High Energy Physics approved by the CERN Council.

The single most critical technology, underlying the LHC and proton colliders beyond the present LHC configuration, is 
superconducting (SC) magnet technology.  Indeed, the present LHC is based on 30 years of development of SC technology 
using NbTi wire. In the LHC NbTi-based magnets are pushed to their limits both in the collider arcs and in the interaction 
regions, designed and built with contributions by US and Japanese national laboratories. 
Delivering the beam brightness required for HL-LHC will pose difficult challenges to the LHC injector chain 
(even after its upgrades), as well as to the LHC storage ring for preserving it. In light of this limitation, the preferred 
route towards increasing luminosity is the reduction of the $\beta^*$ parameter of the beam optics, controlling the beam 
focusing at the interaction point (IP), by means of larger aperture triplet magnets in the interaction region. 
The present design of the HL-LHC interaction region  will require quadrupoles with an aperture of 150~mm and 
peak field in excess of 12~T, beyond the capabilities of the NbTi conductor. Therefore, it relies heavily on the success 
of the advanced Nb$_3$Sn SC technology, developed by an integrated consortium of U.S. national laboratories, the LHC Accelerator 
Research Program (LARP)~\cite{Wanderer:2009zz}. In addition to magnets, many other technologies will be involved in the HL-LHC 
upgrade, namely, crab cavities, advanced collimators, high temperature superconducting links, advanced remote handling, etc. 
LARP is playing a leadership role in this work to maximize the discovery potential of the extensive LHC infrastructure at 
CERN and is instrumental for the success of the HL-LHC upgrade.

{\sl A vigorous continuation of the LARP program leading to U.S. participation in the HL-LHC construction project is crucial 
to the full realization of the potential LHC to deliver discovery physics to the thousands of U.S. researchers who are engaged 
at the LHC.}

In the exploration of the energy frontier with hadron collisions, energy and luminosity play a complementary role. Production 
rates for the signals of interest at the collider may be small either because the mass of the produced objects is large 
or because the coupling strength is small. In the former case, an increase in the beam energy is clearly favorable. 
In contrast, to probe small couplings at smaller masses higher luminosity may be more effective. In reaching toward 
higher energies, magnet technology will retain its pivotal role.
Given the progress in magnet technology and the maturity that Nb$_3$Sn has reached, thanks to the LARP program for the HL-LHC, 
it is legitimate to forecast that Nb$_3$Sn magnets with operating field (that is with 15-20\% margin with respect to quench) 
can reach their limit of 15-16 T in operative condition within the next decade, opening the path towards a collider with 
energy significantly larger than the present LHC, with an exceptional potential for probing the energy frontier.
The first option is a machine housed in the LHC tunnel (HE-LHC). The achievable center-of-mass energy depends on the available dipole 
field strength. An energy of 26~TeV is within the reach of a focused engineering readiness program on Nb$_3$Sn technology, although 
it still requires significant engineering development. The energy reach would become $\sim$33~TeV, if 20~T magnets based on 
futuristic high temperature superconductors (HTS) were practical and affordable and if the constraints imposed by the limited 
space available in the LHC tunnel could be overcome. 
Beyond the confines of the LHC tunnel, a $\sim$100~TeV proton collider would be possible in a larger tunnel. Studies for 
a very large proton collider able to deliver center-of-mass energies of that scale have been conducted over the past two 
decades. In particular, notable was the multi-laboratory VLHC study led by Fermilab in 2001.

The continuation of an integrated, multi-lab program in the US (LARP-like) towards the engineering development of higher field 
magnets, for the HE-LHC in the LHC tunnel or a new collider in big tunnel, needs to be a HEP priority. Such a program 
of innovative engineering, closely coordinated with CERN, would establish the limits of the Nb$_3$Sn technology, 
investigate new conductor materials and re-assess our present concept for managing the enormous stresses produced by 
such high magnetic fields. 
The experience from RHIC, SSC and LHC indicates that the dipoles account for about half of the total collider cost. 
Therefore, magnet technology is an area of critical investment and should represent a major focus of the R\&D process, 
as discussed extensively in this report. 

{\sl  The study group recommends that the US becomes a leading contributor to the design study of a high energy hadron collider.}

With the renewed interest in a $\sim$100~TeV scale collider, also  in the U.S.~\cite{Bhat:2013epd}, the study group recommends 
the participation in the design study for a collider in a large tunnel, which CERN is now organizing.  The construction of a new 
tunnel relieves the pressure from the achievable dipole field strength, since this can be traded for the tunnel circumference. 
The target collision energy is 100 TeV for 20~T dipoles in an 80~km tunnel. However, an 100~km tunnel would provide the 
same collision energy of 100~TeV with reduced field of 15-16~T, reachable with Nb$_3$Sn technology that relies on a much more 
mature and less expensive conductor than HTS. The 100~TeV collider study will inform directions for expanded U.S. technology reach 
and help guide the long-term roadmap of high energy physics capabilities. Areas of particular value to the U.S. R\&D portfolio 
include studies of beam dynamics, magnets, vacuum systems, machine protection and global layout optimization. We note that a large 
tunnel may also open up the possibility of an $e^+e^-$ storage ring~\cite{Sen:2001gr,Koratzinos:2013ncw}, of interest as a $Z$, $W$ 
and Higgs factory if the International Linear Collider is not built.

Whether in the LHC tunnel or in a new, larger tunnel, a collider with energy beyond the LHC will have to deal with an 
additional challenge, the emission of at least 20 times greater beam synchrotron radiation per meter, than the LHC 
at its nominal parameters. The beam pipe and beam screen will have to absorb that radiation. Although synchrotron 
radiation is very beneficial for beam stabilization and will make the higher energy collider the first hadron machine 
dominated by synchrotron radiation damping, the power dissipated in synchrotron radiation must be removed at cryogenic 
temperatures. In the LHC it is removed at 5-10~K. Merely relying on a solution similar to that adopted at the LHC with a beam 
screen at 10~K would be a heavy burden for the cryogenics. For the HE-LHC, a solution based on a beam screen at 40-60~K has 
been envisaged and has no major drawback, although careful design, engineering and prototyping need to be developed to prove 
this solution. For a $\sim$100~TeV machine, handling the synchrotron radiation will become a major challenge.

{\sl Focused engineering development is no substitute for innovative R\&D.} 

In addition to the focused program of engineering development, the separate programs of long range research in magnet 
materials and structures need to be vigourosly continued. It was these programs which have provided the intellectual 
and infrastructure base for the success of LARP.  
For the long range future of high energy physics using $pp$ collisions, advances in magnets of operating 
fields beyond 15-16~T may be needed.  In practice, the decision on the SC technology will rest on a tradeoff between tunnel 
cost, which scales roughly inversely with dipole field, and the cost of the dipole magnets that scales roughly linearly with 
dipole field for long magnets. The crossover in cost cannot be decided {\it ab initio}, as the tunnel cost will depend strongly 
on the geology of the chosen site~\cite{osborne}. 
A new generation of magnets using high temperature superconductors will require new engineering materials with small 
filament size and available in multi-kilometer piece lengths.  Advanced magnets may offer greater temperature margin 
against quenches due to stray radiation lost from the beam. Higher field magnets will require proven stress management 
techniques, exquisitely sensitive magnet protection schemes, and perhaps novel structural materials.
Beam dynamics and other effects of marginal (as compared with electron colliders) synchrotron radiation damping must be 
understood.  As the energy stored in the machine reaches several GigaJoules, control of tenuous beam halos becomes a 
pressing issue. Likewise machine protection from accidental beam loss and the design of special beam abort dumps becomes 
a difficult challenge that requires innovative approaches. Other issues that should not be neglected in a long range R\&D 
program include beam physics of the injection chain, noise and ground motion effects, and the design and technology 
options for the configuration of the interaction regions. Advanced instrumentation provides us with detectors which are faster, 
thinner and have higher segmentation. Their application to imaging for beam monitoring and diagnostics should be carefully 
considered. These areas have all substantial implications for lepton colliders and the intensity frontier. In particular, 
the high-field magnet development is key to the muon accelerator program. 
High energy hadron collisions have produced, and will produce over the next decade or more, some of the most fruitful collider 
physics research. In order to maximize the potential for US HEP, this collider program requires a long-term in-depth 
relation of US laboratories and universities with CERN.

\section{The physics landscape}
\label{sec:phys}

The observation of the Higgs boson at the LHC opens a new era for
particle physics. Its measured properties are consistent, within the
current uncertainties, with those of the Standard Model (SM) Higgs
boson. This gives a remarkable confirmation of the
theoretical setup, formulated over forty years ago, to explain the
otherwise inconsistent co-existence of a gauge theory for electroweak
(EW) interactions, with the masses of the gauge bosons, quarks and charged 
leptons. On the other hand, the complete lack of evidence for new physics makes 
the understanding of the naturalness of the EW scale more concrete and urgent. 
The smallness of the EW scale compared to the Planck scale requires, in the context 
of the SM, an incredible amount of fine tuning.  The existence of new physics beyond
the SM (BSM) is also needed in view of the SM inadequacy to explain intriguing
phenomena like dark matter, the matter-antimatter asymmetry and neutrino 
masses.

After the discovery of the Higgs boson, understanding what is the real
origin of EW symmetry breaking (EWSB) becomes the next key challenge 
for collider physics. This challenge can be expressed in terms of two
questions: up to which level of precision does the Higgs boson behave
as predicted by the SM? Where are the new particles that should
solve the EW naturalness problem and, possibly, offer some insight in
the origin of dark matter, the matter-antimatter asymmetry, and 
neutrino masses?  The first question allows us to define concrete
deliverables for the LHC and future colliders. We know there is a Higgs 
boson at a mass of about 125~GeV and we can thus analyze in great detail 
the prospects for more precise measurements of all of its properties at 
the various facilities.  The second question does not come with the
guarantee of concrete discovery deliverables, but its relevance is powerful 
enough to justify pursuing the search efforts as ambitiously as possible.

The approved LHC programme, its future upgrade towards higher
luminosities, and the study for a new hadron collider delivering
collisions at significantly higher energy, are all essential
components of this endeavor. 
High-luminosity offers the potential to increase the precision of
several key measurements, to uncover rare processes, and to guide and
validate the progress in theoretical modeling, thus reducing the
systematic uncertainties in the interpretation of the data. The
extended lever arm afforded by an higher collision energy increases the
potential to directly probe greater and greater mass scales for BSM
processes, and gives access to the TeV energy scale, which is the
natural domain to test the dynamics of EWSB (e.g. high-mass $WW\to WW$
and $WW\to HH$ scattering, triple and quartic gauge and Higgs
couplings).  Detailed studies of the realistic performance in the
measurements of Higgs couplings, self-coupling and $WW$ scattering at
energies above 14~TeV do not yet exist and will be required to 
precisely assess their accuracy and potential.
More in general, higher statistics, whether coming from
higher luminosity or from increased cross sections at higher energies,
open new opportunities for both searches and measurements, enabling
analysis strategies where tighter cuts can greatly improve the signal
purity and/or reduce the theoretical uncertainties. The long
history of the Tevatron gives clear evidence of the immense progress
that can be made, relative to naive earlier estimates, after the
accumulation of bigger amounts of data and of more experience with
their analysis.

The lack of a guarantee that {\it any} future facility will directly
observe new particles greatly strengthens the relevance of a broad
programme of very precise measurements. The exploration of the Higgs
sector provides us with a benchmark for the assessment of the physics
potential of future facilities, including hadron colliders at the
energy frontier. Several directions should be pursued, including the
high-precision study of the couplings already observed, the search 
of rare (e.g. $H\to \mu^+\mu^-$) or forbidden Higgs
decays and the study of the dynamics of Higgs interactions, including
Higgs-pair production processes, sensitive to the Higgs
self-coupling and to possible anomalous interactions.

What really defines the Higgs boson is its role in breaking the
EW symmetry; in particular, its coupling to the longitudinal
polarization of $W$ and $Z$ bosons and the resulting unitarization of
high-energy $WW$ scattering.  This is a key element in the study of
EWSB. The theoretical description of $WW$ scattering requires the
exchange of the Higgs boson, or of some other new particle, in order
to tame the otherwise unphysical rate growth at energies around the
TeV and above. Verifying the details of this process is essential to
learn more on whether the Higgs boson is indeed a fundamental particle or,
as postulated in some theories, a composite object, something that
could also manifest itself with the appearance of new resonances in
the TeV range. As it has been known and documented for many years, these
studies require the highest possible energies. For example, in a class
of composite-Higgs models, deviations
from the SM behavior of the $WW$ scattering cross section, 
due to anomalous Higgs interactions,
scale like  $\xi^2 (E_{CM}(WW) /
600~{\mathrm{GeV}})^4$, where $\xi$ is related to the scale of new
physics\footnote{$\xi=v^2/f^2$, where $v=246$~GeV is the EWSB scale, and
  $f$ is the compositeness scale.}, and determines also a change of order
$\xi$, w.r.t. the SM prediction, in BR($H\to WW^{*}$). 
It takes center-of-mass energies of the $WW$
system well above 1~TeV to have sensitivity to these deviations. At
14~TeV and 300~fb$^{-1}$, the statistics of events with $E_{CM}(WW) >$ 1~TeV is
limited and experiments will only be sensitive to values $\xi\gsim 0.5$. 
This sensitivity improves to $\xi$ values of 
$\cal{O}$(0.10) and $\cal{O}$(0.01) for $pp$ collisions at 30 and
100~TeV, respectively. The design energy of the SSC, 40~TeV, was
chosen to optimize the reach and precision of these measurements. The
need to perform these studies today is even stronger than it was
in the days of the SSC planning, due to the observation of a light
Higgs particle.  As already mentioned, no realistic assessment of the potential 
of possible future experiments for these measurements is currently available. The
issues to be considered include the geometrical acceptance of
forward/backward jets and the ability to reconstruct them in presence
of large pileup of underlying events as discussed below.



\begin{table}
\begin{center}
\caption{Evolution of the cross sections for different Higgs production 
  processes in $pp$ collisions with center-of-mass energy  The cross sections 
  at $\sqrt{S}=14$~TeV are given in the second column, and the
  ratios $R(E)=\sigma(E~\mathrm{TeV})/\sigma(E=14~\mathrm{TeV})$ in
  the following columns. All rates assume $M_H=125$~GeV and SM
  couplings.}
\label{tab:Hrates}
\begin{tabular}{|l|c|c|c|c|c|c|}
 \hline
 Process & $\sigma$(14 TeV) & R(33) & R(40) & R(60) & R(80) & R(100)
 \\
\hline
$gg\to  H$ & 50.4 pb & 3.5 & 4.6 & 7.8 & 11&  15
 \\
\hline
$qq\to qq H$ &  4.40 pb & 3.8 & 5.2 & 9.3 & 14 & 19
 \\
\hline
$q\bar{q} \to WH$ & 1.63 pb & 2.9 & 3.6 & 5.7 & 7.7 & 10
 \\
\hline
$q\bar{q} \to ZH$ &  0.90 pb & 3.3 & 4.2 & 6.8 & 10 & 13
 \\
\hline
$pp\to HH$ & 33.8 fb & 6.1 & 8.8 & 18 & 29 & 42
\\ \hline
$pp\to ttH$ & 0.62 pb & 7.3 & 11 & 24 & 41 & 61
\\ \hline
\end{tabular}
\end{center}
\end{table}
Table~\ref{tab:Hrates}~\cite{HXS-SG} summarizes the increase in rate for 
several Higgs production channels in $pp$ collisions, as a function of the 
beam energy, covering the range of possibilities being considered in this
document. Final states with the largest invariant mass (like $ttH$ and $HH$) 
benefit the most from the energy increase. This benefit is further enhanced 
when we consider the fraction of events passing the typical analysis cuts 
imposed to improve signal separation or reduce systematic uncertainties. 
For example, in the case of $ttH$ production, requiring the top quarks to
have a transverse momentum above 500~GeV would increase the rates by a
factor of 16 (250) at 33~TeV (100~TeV), instead of the factor of 7
(60) increase for the fully inclusive $ttH$ rate.

In many BSM scenarios for EWSB, the Higgs boson is accompanied by
several new particles, with masses in the range of hundred(s) GeV up to
possibly several TeV. These could be other Higgs-like scalar states,
or heavier partners of the $W$ and $Z$ gauge bosons and of the top and
bottom quarks, or, as in the case of supersymmetry, a complete
replica of the SM particle spectrum, where each known particle has a 
partner, with a different spin quantum number. 
Any deviation of the measured Higgs properties from the SM expectation would 
imply the existence of at least some of these particles, and viceversa. 
The mass limits on such new particles derived so far from the LHC still leave 
ample room for their discovery in the 14~TeV data. 
However, any discovery at 14~TeV will require a follow-up phase
of precision measurements, to understand the origin of the newly
observed phenomena. For example, while new $Z^\prime$ gauge bosons,
signaling the existence of new weak interactions, can be discovered
with 300~fb$^{-1}$ up to 4-5~TeV, the full HL-LHC luminosity will be needed
to determine their properties if their mass is above 2.5~TeV. Existing
studies also show that a tenfold increase in the LHC integrated
luminosity will extend the discovery reach for new particles by 30-50\%.

\begin{figure}
\begin{center}
\includegraphics[width=0.75\textwidth,clip]{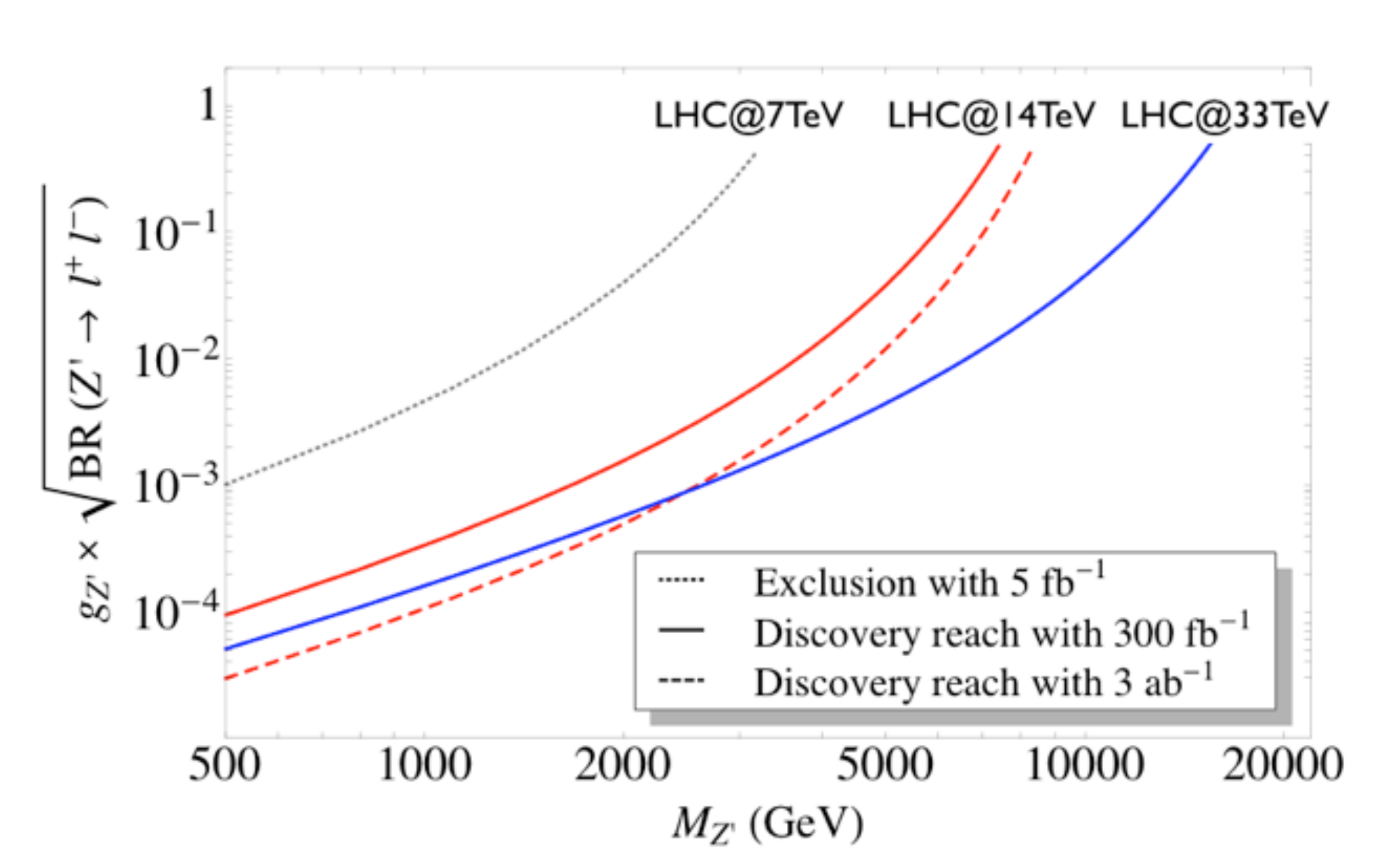}
\caption{\label{fig:Zprime} 
Discovery potential for new $Z^\prime$ gauge bosons decaying to lepton pairs,
as a function of their mass. The reach is expressed in terms of their
coupling strength times the leptonic branching fraction.}
\end{center}
\end{figure}

In the study of the energy frontier with hadron collisions energy and
luminosity play a complementary role.  Production rates for the
signals of interest may be small either because the
mass of the produced objects is large or because the coupling strength
is small. In the former case, an increase in the beam energy is
clearly favorable, while to probe small couplings at smaller masses
higher luminosity may be more effective.  For a signal at a fixed mass
scale $M$, the cross section $\sigma(M,g) \propto g^2/M^2 \times
L(x=M/\sqrt{S})$ grows with the hadronic center of mass energy
$\sqrt{S}$, since the {\it partonic} luminosity $L(x)$ grows at least
like $\log(1/x)$. An increase in accelerator energy does not need, in
this case, to be accompanied by an increase in luminosity. On the
other hand, to scale the discovery mass reach $M$ with the beam energy
means keeping $x=M/\sqrt{S}$ fixed. In this case the cross section
scales like $\sigma(M,g) \propto g^2/{S} \times L(x)/x \propto 1/S$
and the collider luminosity must grow as the square of the energy.

Depending on the mass and couplings of these new particles
the most effective way to increase their statistics could be either higher
luminosity or higher energy. This is illustrated in a simple concrete
case in Fig.\ref{fig:Zprime}, which represents the LHC discovery
potential for new $Z^\prime$ gauge bosons decaying to lepton pairs, in different
energy and integrated luminosity conditions.  The reach is expressed
by the product of their coupling strength, $g$, and the square root
of the leptonic decay branching fraction.  The dashed line
corresponds to the result after 3000~fb$^{-1}$ at 14 TeV, the solid
blue line to 300~fb$^{-1}$ at 33~TeV. We notice that for masses below
$\sim 2.5$~TeV the factor of ten higher luminosity at 14~TeV leads to
a better discovery reach (or, in the case of a previous discovery,
leads to higher statistics and better precision in the measurement of
the $Z^\prime$ properties). On the contrary, if the $Z^\prime$ is heavier than
$\sim 2.5$~TeV, the increase is energy is more effective. The figure
also shows that the run at 7~TeV has already excluded the existence of
$Z^\prime$ bosons up to 2.5~TeV, at least for some range of their
couplings. This means that, at least for some models, a new particle 
discovered at 14~TeV would be sufficiently heavy that its precision
studies would greatly benefit from the energy upgrade, even after the
completion of an extensive high luminosity LHC phase. Similar
reasoning applies to other BSM new particles. A higher energy $pp$ collider 
is therefore a powerful tool to extend and improve the precision studies
of the Higgs boson and other phenomena to be uncovered during the
nominal and high-luminosity LHC runs at 14~TeV, as well as to open the
way for the exploration of a new energy range, unattainable by any of
the other current proposals for new high-energy facilities.

The exploration of physics beyond the SM at the high energy frontier will 
unavoidably require going towards hadron collisions at higher energies. 
The current limits obtained by LHC at 8~TeV already point in a natural way 
to higher energies as the best way to perform quantitative studies 
of possible discoveries made at the LHC at 14~TeV. Even in the absence of such 
discoveries, the exploratory potential of a high energy $pp$ collider 
will provide the only opportunity to shed more light on the origin on EW 
symmetry breaking and of the hierarchy problem at the energy frontier.

\section{Underlying events in high energy pp collisions}
\label{sec:ue}

Underlying events (UE) represent a major challenge for experiments at high-luminosity 
hadron colliders both in terms of occupancy and of precision in the reconstruction of 
the event kinematics. The beam parameters for the LHC high-luminosity upgrade,
discussed below, are chosen to mitigate the number of underlying events and keep 
them within manageable limits. The evaluation of the rate and characteristics of 
minimum bias events as a function of the center-of-mass energy is therefore important 
for guiding the choice of beam parameters from 14~TeV towards higher energies~\cite{Skands:2013asa}.
\begin{figure}[ht!]
\centering
\begin{tabular}{c}
 \includegraphics*[scale=0.40]{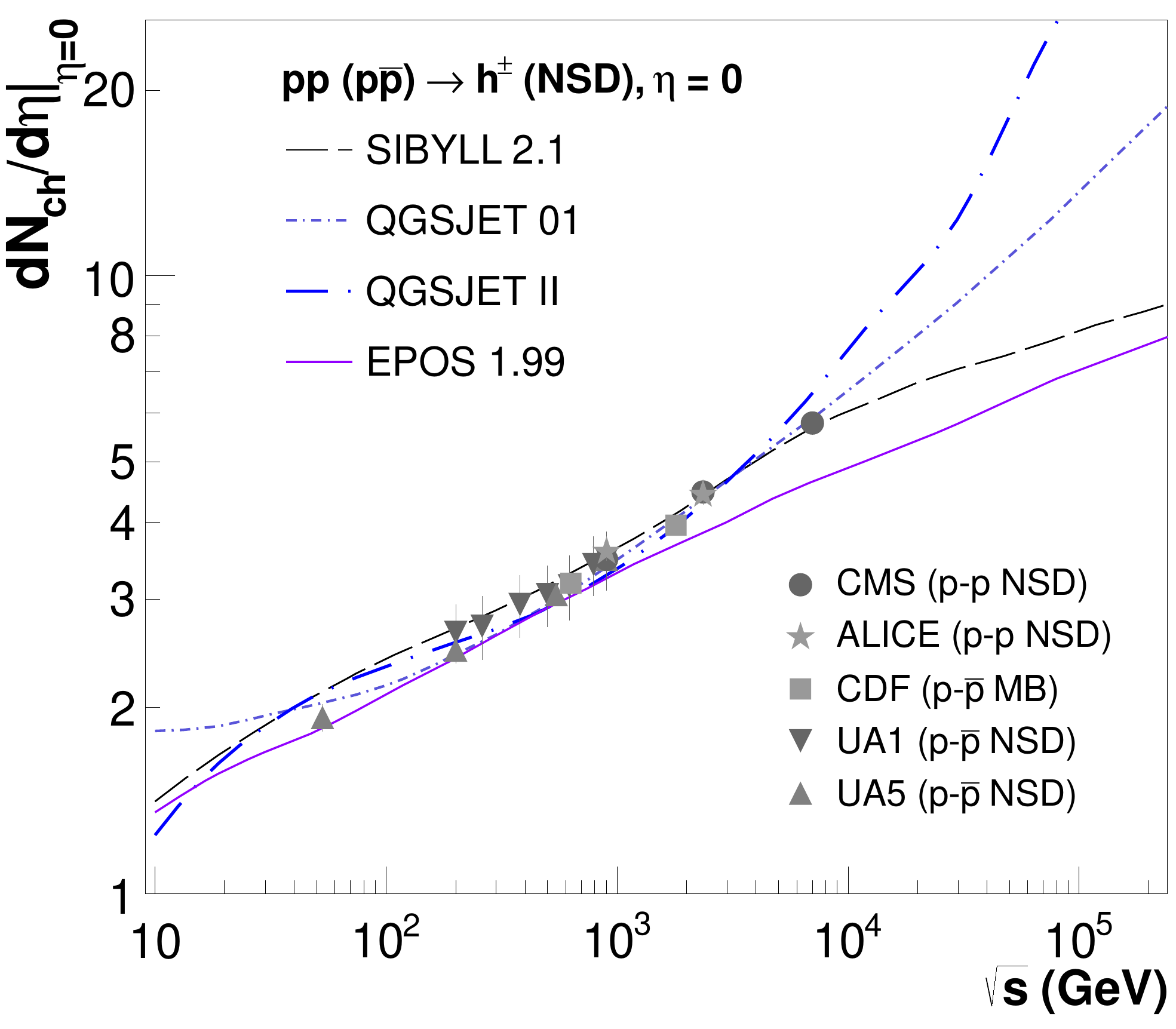}
\end{tabular}
\caption{Scaling of the central charged multiplicity for the
  \sib, \qgs, and \epos models compared to collider data for NSD
  events (from~\cite{d'Enterria:2011kw}, courtesy of the author). 
\label{fig:dndeta}
}
\end{figure}

The total cross section of minimum bias events, can be obtained with a simple 
Donnachie-Landshoff fit with $\epsilon\sim 0.08$~\cite{Donnachie:1992ny}, 
similarly to the scaling ans\"atze made in \py~\cite{Schuler:1993wr,Sjostrand:2006za}. 
The ALICE measurements of the inelastic and single-diffractive cross 
sections~\cite{Abelev:2012sea} do not show any significant deviations from this 
ansatz over the kinematic range of their measurements.
The extrapolations yield an inelastic cross section growing from $\sim$70~mb at 
7~TeV to $\sim$90~mb at 30~TeV and $\sim$105~mb at 100~TeV. The diffractive components 
increase by only a few mb relative to the LHC values. These collisions can be characterized 
in terms of their track multiplicity and associated energy deposition. The extrapolations 
of central charged-track densities in so-called non-single-diffractive events in pomeron-based 
models are shown in the left-hand panel of \figRef{fig:dndeta} (from \cite{d'Enterria:2011kw}).  
\begin{figure}[hb!]
\centering
\begin{tabular}{rl}
&\hspace*{0.5cm}\includegraphics*[scale=0.50]{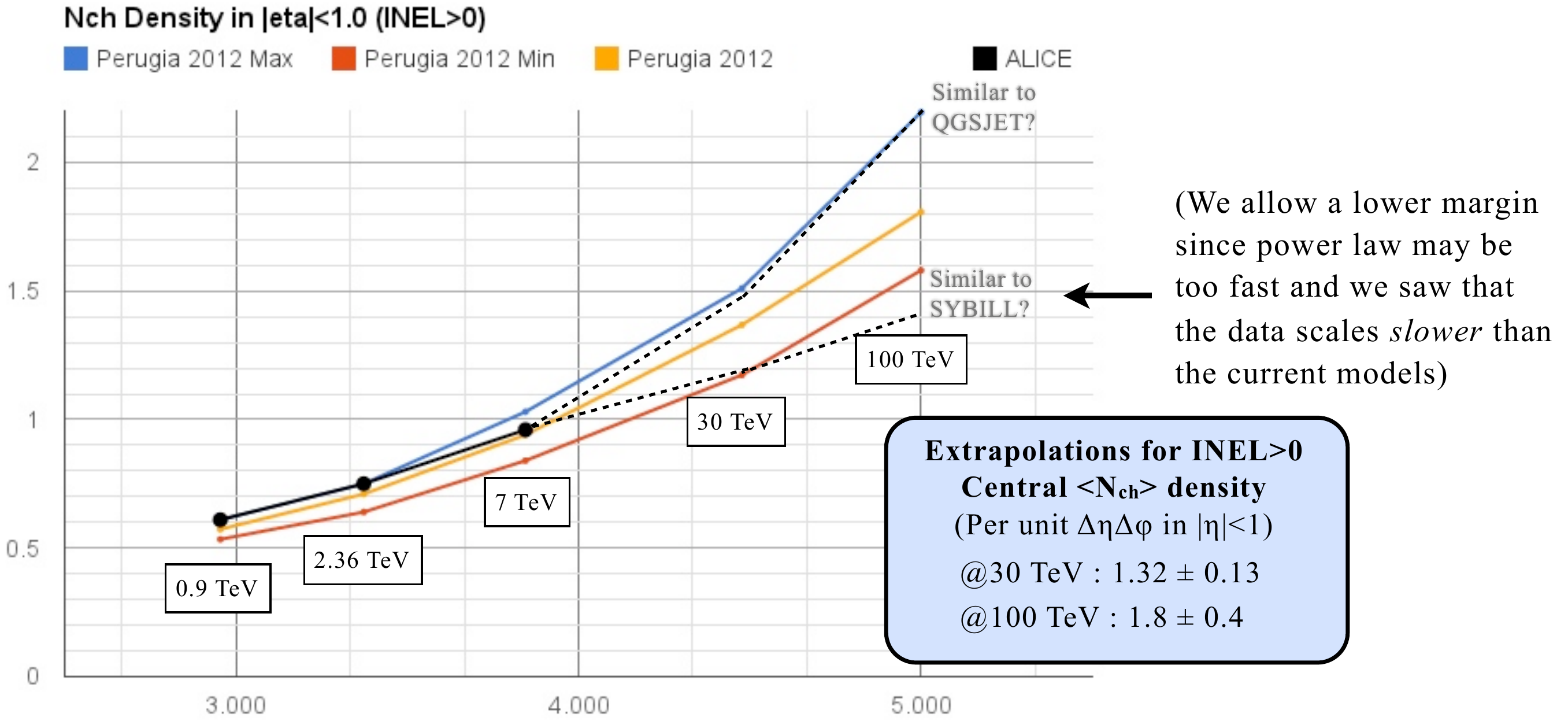}\\
& \small\hspace*{4.6cm}$\log_{10}(E_\mrm{CM}/\mrm{GeV})$\\[3mm]
&\hspace*{0.05cm}\includegraphics*[scale=0.53]{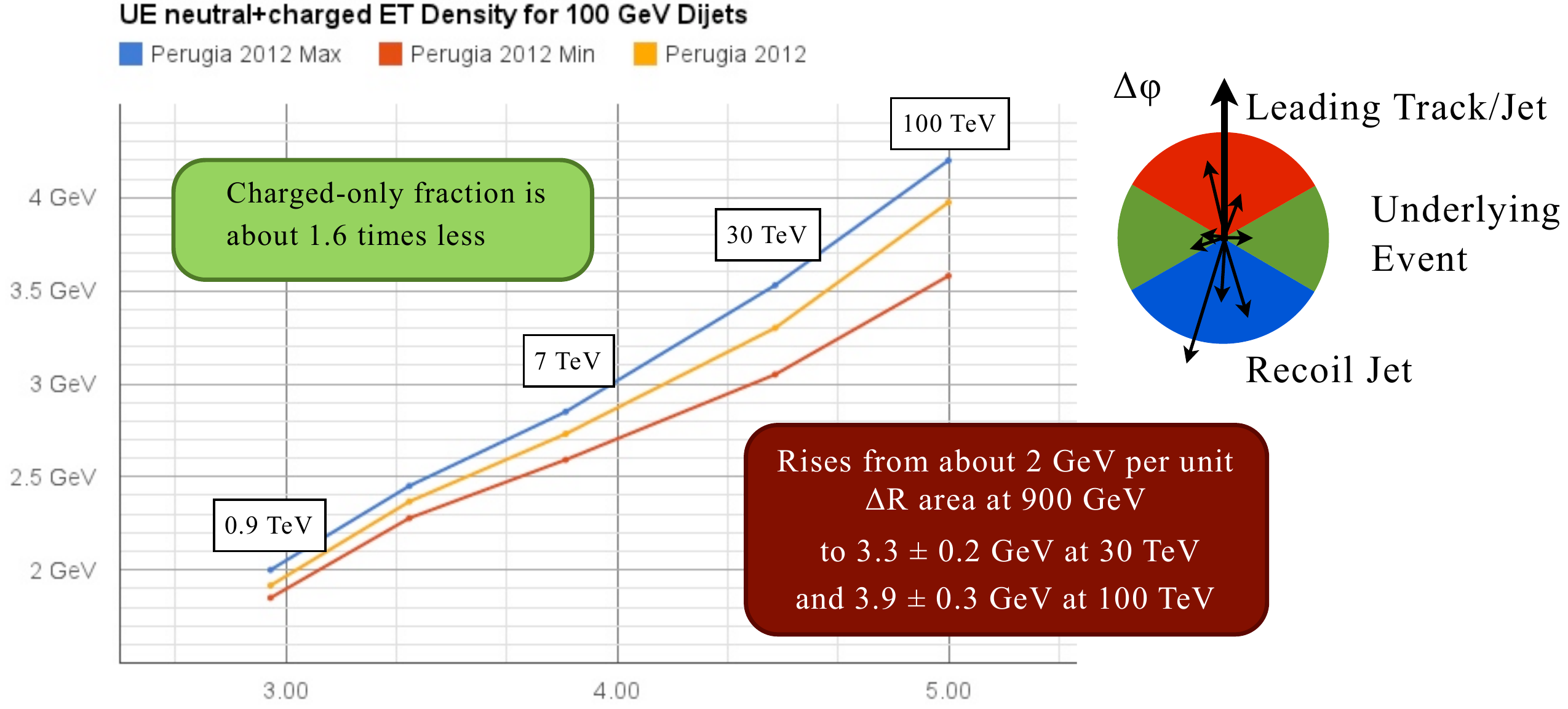}\\
& \small\hspace*{4.6cm}$\log_{10}(E_\mrm{CM}/\mrm{GeV})$
\end{tabular}
\caption{Extrapolations of the central charged particle density for
  INEL$>$0 events ({\sl
    Top}), 
and central UE $E_T$ density for 100-GeV di-jet events ({\sl Bottom}). 
\label{fig:extrapolations}}
\end{figure}
Combining the uncertainty variations of various generator tunes, when including only those 
compatible with the scaling observed at the LHC, yields an estimated 
central charged-track density per unit $\Delta\eta\Delta\phi$ of $1.32\pm0.13$ at 30~TeV, 
and $1.8\pm0.4$ at 100~TeV, for inelastic events with at least one track inside the 
$|\eta|<1$ acceptance, which represent $\simeq$85\% of all inelastic collisions at 30 -- 100~TeV, 
compared to about 80\% at LHC energies.

An important quantity for jet energy scale calibrations is the amount of 
transverse energy deposited in the detector, per unit 
$\Delta R^2=\Delta\eta\times\Delta\phi$, per inelastic collision.
In the central region of the detector, the Perugia 2012 models~\cite{Skands:2010ak} 
are in good agreement with the ATLAS measurements at 7~TeV~\cite{Aad:2012mfa,Karneyeu:2013aha}, 
while the activity in the forward region appears to be
underestimated~\cite{Chatrchyan:2011wm,Aad:2012mfa,Aspell:2012ux,Karneyeu:2013aha}.  
Extrapolations lead to an estimated $(1.25\pm0.2)\,\mrm{GeV}$ of transverse energy 
deposited per unit $\Delta R^2$ in the central region of the detector at 30~TeV, 
growing to $(1.9\pm0.35)\,\mrm{GeV}$ at 100~TeV. 

The last quantity we consider is the activity in the underlying event. 
The most important UE observable is the summed $p_\perp$ density in the 
so-called ``transverse'' region, defined as the wedge $60-120^\circ$ away 
in azimuth from a hard trigger jet. For $p_{\perp}^\mrm{jet}$ values above 
5 -- 10 GeV, this distribution is effectively flat, i.e. to first approximation 
it is independent of the jet $p_\perp$. It does, however, depend significantly 
on the center-of-mass energy of the $pp$ collision, a feature which places 
strong constraints on the scaling of the $p_{\perp0}$ scale of MPI models. 
Given the good agreement between the Perugia 2012 models and Tevatron
and LHC UE measurements~\cite{Karneyeu:2013aha}, we estimate the  
$E_T$ (neutral+charged) density in the transverse region (inside
$|\eta|<2.5$), for a reference case of 100~GeV di-jets in the bottom
pane of \figRef{fig:extrapolations}. Starting 
from an average of about $2\,\mrm{GeV}$ per
$\Delta R^2$ at 900 GeV, the density rises to $(3.3\pm 0.2)\,\mrm{GeV}$ 
at 30 TeV and to $(3.9\pm 0.3)\,\mrm{GeV}$ at 100 TeV, while the charged-only 
fraction should be lower by a factor $\sim$1.6. 

\section{LHC luminosity upgrade}
\label{sec:lhc}



\begin{figure}[h!]
        \centering
        \includegraphics[width=0.75\textwidth]{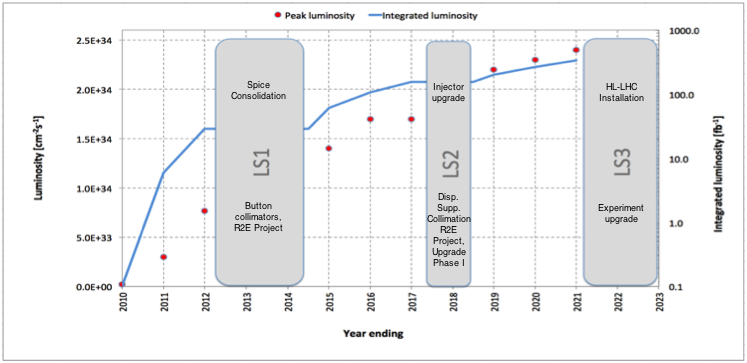}
        \caption{LHC baseline plan for the next ten years. The first long shutdown (LS) 2013-14 is to allow 
design parameters of beam energy and luminosity. The second LS, 2018, is to increase beam intensity and reliability 
as well as to upgrade the LHC Injectors. A forecast of the peak luminosity and integrated luminosity is given.}
        \label{fig:LHC-luminosity}
 \end{figure}

The full exploitation of the LHC is the highest priority of the energy frontier 
collider program. LHC is expected to restart in Spring 2015 at center-of-mass energy of 
13-14 TeV and that it  will reach the design luminosity of 10$^{34}$ cm$^{-2}$s$^{-1}$ 
during 2015. This peak value should give a total integrated luminosity over 
one year of about 40 fb$^{-1}$. In the period 2015-2020, the LHC will gradually increase 
its peak luminosity. Margins have been taken in the design to allow the machine to reach 
about two times the nominal design performance. The baseline program for the next decade 
is schematically shown in Figure~\ref{fig:LHC-luminosity} with the expected evolution of 
the peak and integrated luminosity.

The LHC has provided collisions so far at 8~TeV center-of-mass energy, and with total beam currents of 
about 0.4~A (i.e.\ 70\% of the nominal design value but with only half the nominal number of bunches). 
Once the magnet interconnections have been consolidated and the beam energy limits removed, as well as 
some radiation-to-electronic (R2E) intensity limits mitigated during LS1 phase in 2013-14, the design luminosity 
will hopefully be attained and eventually overcome in 2015. Then, by removing the outstanding beam intensity limits 
in the injector chain and the LHC during additional shutdowns following the LS1, the LHC will 
head toward the so called {\sl ultimate} design luminosity, which is about twice the nominal luminosity, 
i.e., 2$\times$10$^{34}$ cm$^{-2}$s$^{-1}$. This ultimate luminosity performance was planned to be reached 
by increasing the bunch population from 1.15 to 1.7$\times$10$^{11}$ protons, with a bunch spacing of 25~ns 
(beam current increases from 0.58~A to 0.86~A). Transforming this ultimate peak performance into a doubling 
of the annual integrated luminosity will however be very difficult and it is more likely that the delivered 
luminosity  will be around 60-70~fb$^{-1}$/year.

After 2020 some critical components of the accelerator will reach the radiation damage 
limit and other will see their reliability reduced because of vulnerability to radiation, 
wear and high intensity beam operation. Therefore, important consolidation actions are required 
before 2020, just to keep the LHC running with a good availability. Further, the statistical gain 
in running the accelerator without an additional considerable luminosity increase beyond its 
design value will become marginal, therefore  the LHC will need to have a decisive increase of 
its luminosity.

This new phase of the LHC life, named High Luminosity LHC (HL-LHC), has the scope of preparing 
the machine to attain the astonishing threshold of 3000 fb$^{-1}$ during 10-12 years of operation.
The project is now the first priority of Europe, as stated by the Strategy update for High Energy 
Physics group approved by CERN Council in a special session held on May, 30 2013 in Brussels.

\subsection{HL-LHC}

The LHC $pp$ nominal design luminosity for each of the two general purpose experiments 
(ATLAS and CMS) is $L_0$=10$^{34}$ cm$^{-2}$s$^{-1}$. This luminosity is associated with a bunch 
spacing of 25~ns (2808 bunches per beam) and gives an average value of 27~underlying events/crossing. 
The main objectives of HL-LHC are the following:
\begin{enumerate}
\item peak luminosity of 5$\times$10$^{34}$cm$^{-2}$s$^{-1}$ with leveling,
\item integrated luminosity of 250 fb$^{-1}$/year, enabling the goal of 3000 fb$^{-1}$ within twelve 
years after the upgrade. 
\end{enumerate}
This luminosity is about ten times the luminosity reach of the first twelve years of the LHC lifetime. 
The luminosity upgrade provides the HEP community with an unprecedented data sample taken at the 
frontier constituent energy, which will be key to investigate the dynamic of EWSB and answer some of the 
most important open questions in particle physics, as discussed above.

The HL-LHC project is matched to a companion LHC detector upgrade program, to ensure that the detectors 
will keep their outstanding performance while operating with an average of $\sim$140 underlying events. 
The 5$\times$10$^{34}$ cm$^{-2}$s$^{-1}$ value for the luminosity leveling corresponds to  
the nominal 25~ns bunch spacing. For a 50~ns bunch spacing, the leveling value would be half of this value, 
only, with the unavoidable loss of integrated luminosity. The experiments are actually designing their 
upgraded detectors to be capable to sustain a maximum of 190 events/crossing, keeping a 
reasonable margin against shortfalls, including a possible run of the machine at 50~ns should 25~ns become 
too difficult, and for taking into account the pile-up fluctuations around the average value~\cite{CMS:2013xfa}. 

\subsection{Beam parameters}

Experience with the LHC shows that the best set of parameters for actual operation is difficult to predict, before 
we obtain operational experience with the 7~TeV beams. The upgrade studies should therefore 
provide the required HL-LHC performance over a wide range of parameters, and the machine and experiments will find 
the best set of parameters during operation, once the LHC runs at the maximum energy and with above nominal beam 
intensities.

\begin{table}[h!]
\begin{center}
\caption{Parameters of LHC, HL-LHC, HE-LHC, and VHE-LHC. The luminosity given for HL-LHC assumes the use of crab cavities.}
{\begin{tabular}{@{}lcccc@{}} 
\hline
       \textbf{parameter} & \textbf{LHC} & \textbf{HL-LHC}~\cite{Zimmermann:2011za,hl} & \textbf{HE-LHC}~\cite{helhc} & \textbf{VHE-LHC}~\cite{vhe} \\ 
\hline
                c.m.~energy [TeV] & 14 & 14 &  33 & 100 \\  
                circumference $C$ [km] & 26.7 & 26.7 & 26.7 & 80 \\  
                dipole field [T] & 8.33 & 8.33 &  20 & 20  \\ 
                dipole coil aperture [mm] & 56 & 56 &  40 & $\le40$ \\ 
                beam half aperture [cm] & $\sim 2$ & $\sim 2$ &  1.3 & $\le 1.3$ \\ 
                injection energy [TeV] & 0.45 & 0.45 &  $>$1.0 & $>$3.0 \\ 
                no.~of bunches $n_{b}$ & 2808 & 2808 & 2808 & 8420 \\ 
                bunch population  $N_{b}$ [$10^{11}$] & 1.15 & 2.2 & 0.94 & 0.97 \\ 
                init.~transv.~norm.~emit.~[$\mu$m] & 3.75 & 2.5 & 1.38 & 2.15 \\ 
                initial longitudinal emit.~[eVs]& 2.5 & 2.5 &  3.8 & 13.5 \\ 
                no.~IPs contributing to tune shift & 3 & 2  & 2 & 2 \\ 
                max.~total beam-beam tune shift & 0.01 & 0.015 &  0.01 & 0.01 \\ 
                beam circulating current [A] & 0.584 & 1.12 & 0.478 & 0.492 \\ 
                rms bunch length  [cm] & 7.55 & 7.55 & 7.55 & 7.55 \\ 
                IP beta function [m] & 0.55 & 0.15 (min.) & 0.35 & 1.1  \\ 
                rms IP spot size [$\mu$m] & 16.7 & 7.1 (min.) &  5.2 & 6.7  \\ 
                full crossing angle [$\mu$rad] & 285 
                    & 590 & 185 & 72  \\ 
                stored beam energy [MJ] & 362 & 694  & 701 & 6610 \\ 
                SR power per ring [kW] & 3.6 & 7.3 & 96.2 & 2900 \\ 
                arc SR heat load [W/m/aperture] & 0.17 & 0.33 & 4.35 & 43.3 \\ 
                energy loss per turn [keV] & 6.7 & 6.7 & 201 & 5857 \\ 
                critical photon energy [eV] & 44 & 44  &575 & 5474 \\ 
                photon flux [$10^{17}$/m/s] & 1.0 & 2.0 & 1.9  & 2.0 \\
                longit.~SR emit.~damping time [h] & 12.9 & 12.9 & 1.0 & 0.32 \\   
                horiz.~SR emit.~damping time [h] & 25.8 & 25.8 & 2.0 & 0.64  \\ 
                init.~longit.~IBS emit.~rise time [h] & 57 & 23.3 & 40 & 396 \\  
                init.~horiz.~IBS emit.~rise time [h] & 103 & 10.4 & 20 & 157 \\  
                peak events per crossing & 27 & 135 (lev.) & 147 & 171 \\  
                total/inelastic cross section [mb] & \multicolumn{2}{c}{111 / 85} & 129 / 93 & 153 / 108 \\ 
                peak luminosity [$10^{34}$~cm$^{-2}$s$^{-1}$] & 1.0 & 5.0 & 5.0 & 5.0 \\ 
                beam lifetime due to burn off [h] & 45 & 15.4 & 5.7 & 14.8 \\ 
                optimum run time [h] & 15.2 & 10.2 &  5.8 & 10.7 \\ 
                opt.~av.~int.~luminosity / day [fb$^{-1}$] & 0.47 & 2.8 & 1.4 & 2.1 \\ \hline
   \end{tabular}
}
\end{center}
\label{tab:lhc} 
\end{table}

The (instantaneous) luminosity L can be expressed as:
 \begin{equation}
 L= \gamma  \frac{n_b N^2 f_{rev}}{4\pi \beta^*  \epsilon_n } R;  
 \hspace{0.25cm}  R=\frac{1}{\sqrt{1+ \big( \frac{\theta_c  \sigma_z}{2\sigma_x} \big)^2}} 
 \label{eq:luminosity}
 \end{equation}
where $\gamma$  is the proton beam energy in unit of rest mass, $n_b$ is the number of bunches in the machine: 1380 for 50 ns spacing 
and 2808 for 25 ns, $N$ is the bunch population. $N_{nominal, 25 ns}= 1.15\times 10^{11} p$ ($\rightarrow 0.58 A$ of beam current at $2808$  
bunches), $f_{rev}$ is the revolution frequency ($11.2$ kHz), $\beta^*$ is the beam beta function (focal length) at the collision point 
(nominal design $0.55$ m), $\epsilon_n$ is the transverse normalized emittance (nominal design: $3.75$ $\mu$m), $R$ is a luminosity 
geometrical reduction factor depending on the so-called Piwinski angle 
($0.85$ at $0.55$ m of $\beta^*$, down to $0.5$ at $0.25$ m), $\theta_c$ is the full crossing angle between 
colliding beam ($285$  $\mu$rad as nominal design), $\sigma_x$, $\sigma_z$ are the transverse and longitudinal r.m.s. size, 
respectively ($16.7$ $\mu$m and $7.55$ cm).
Table~\ref{tab:lhc} lists the main parameters of the accelerator configurations considered in this report. The parameters for the 
HL-LHC are given in the third column~\cite{Zimmermann:2011za,hl}. The 25~ns bunch spacing is the nominal operating target. However, 
a scheme with 50~ns is also being considered, as fall-back solution. In order to reach the goal of 250 fb$^{-1}$/year with 
160 days dedicated to proton physics, the efficiency must be 60\%. A big leap forward is required by increasing the 
availability  and the turnaround time, i.e.\ the time from end of physics to next start of physics.  

The nominal total beam current of 1.12~A (see Table ~\ref{tab:lhc}) is a difficult target to attain. It represents a hard limit for the LHC 
since it affects many systems, such as RF power systems and RF cavities, collimation, cryogenics, kicker magnets, 
vacuum system, beam diagnostics, in a direct way, and several others, like quench detection system of the SC magnets 
and virtually all controllers, in an indirect way, due to an increase of the R2E events. 
Transverse emittance is assumed to be very low also in view of the already better than the design value results 
during the first LHC running. However, getting the beam brightness, $N_p/\epsilon_n$, required for HL-LHC is a 
very difficult challenge for the injector chain even after upgrades as well as for LHC to preserve it. 

In view of these limitations, the classical route availabe towards the luminosity increase is the reduction of strength of the 
beam focusing at the IP, $\beta^*$, by means of triplet magnets which have larger aperture for a given gradient or are longer 
and larger aperture low-$\beta$ triplet quadrupoles with a reduced gradient. A reduction in the $\beta^*$ value implies 
an increase of beam sizes over the whole matching section. 
Therefore the reduction in $\beta^*$ implies not only larger triplet magnets but also larger 
separation/recombination dipoles and larger and/or modified matching section quadrupoles. A previous study showed 
that a practical limit in the LHC arises around $\beta^*$ = 30-40~cm However, the novel Achromatic Telescopic Squeeze (ATS) scheme 
recently proposed would allow to overcome these limitations in the LHC matching section. 
Thanks to the Nb$_3$Sn quadrupoles of larger aperture and higher field a $\beta^*$ value of 15~cm or even 10~cm can now be envisaged 
and flat optics with a $\beta^*$ as low as 5~cm in the plane perpendicular to the crossing plane are possible. In particular, a 
value of $\beta^*$=10~cm has been recently attained in an LHC machine development run dedicated to test the ATS principle. 
This offers a margin compared to the parameters given in Table~\ref{tab:lhc}.

\begin{figure}[tp]
\centering
\hspace*{0.5cm}\includegraphics[width=0.5\textwidth,clip]{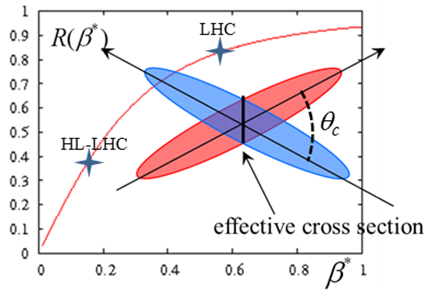}\\
\caption{Geometrical reduction factor of luminosity, $R$, vs. $\beta^*$ with the two operating points for the nominal 
LHC and HL-LHC beam parameters indicated by the crosses.}. 
\label{fig:bextrapolations}
\end{figure}

In order to be compatible with such small $\beta^*$ values, the aperture of the low-$\beta^*$ quadrupole magnets needs to be doubled, 
which causes a peak field 50\% higher than that of the present LHC triplet magnets and requires a new superconducting magnet technology 
based on Nb$_3$Sn, as discussed in the next section.
The drawback of the very small $\beta^*$ values is that they require a larger crossing angle, which entails in turn 
a reduction of the luminosity via the geometrical factor of beam overlap, $R$, defined in Eq~\ref{eq:luminosity}, 
compared to the LHC present conditions (see Figure~\ref{fig:bextrapolations}). The reduction of the beam separation 
at the parasitic encounters and the mitigation of the beam-beam effects are under study but not yet fully demonstrated. 
In the HL-LHC design, the reduction in the geometrical factor $R$  is compensated with the use of crab cavities, 
which rotate the beams before collisions to maximize their overlap~\cite{DeMaria:2011zb}. 
Their development is discussed in the next section.

\section{Technology aspects}
\label{sec:tech}




The LHC is based on 30 years of development in the domain of superconducting (SC) magnet technologies. 
NbTi based magnets are pushed to their limits: very compact two-in-one dipole magnets provide 8.3~T operating 
field by using superfluid helium cooling.

\begin{figure}
        \centering
        \includegraphics[width=0.75\textwidth]{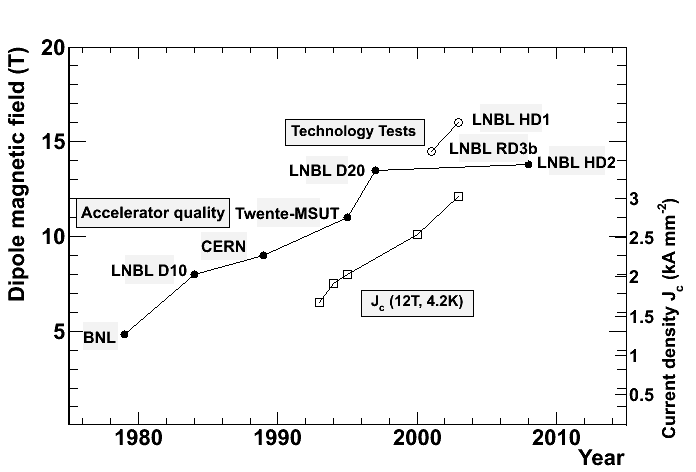}
        \caption{Progress in superconducting dipole magnet performance and in the increase in $J_c(12T, 4.2K)$ of Nb$_3$Sn over 
                 the last 20+ years. The RD3 and HD1 refer to simple race track prototypes for technology tests.}
        \label{fig:progress1}
 \end{figure}

\begin{figure}
        \centering
        \includegraphics[width=0.60\textwidth]{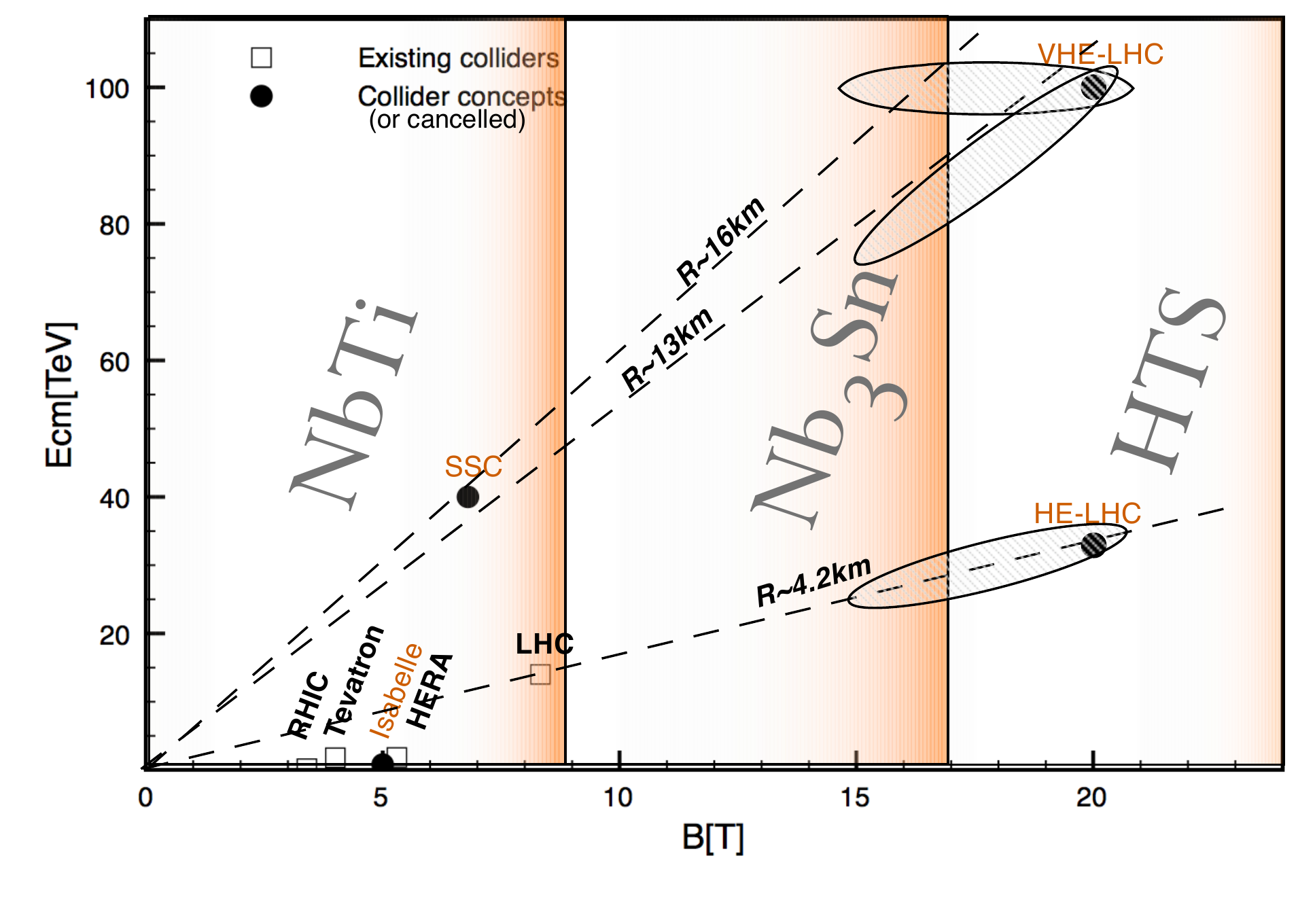}
        \caption{Role of the superconductor in the energy reach of hadron colliders as a function of field and radius. 
	The ellipses show the tradeoff in attainable energy and ring radius associated with the use of HTS materials 
	compared to Nb$_3$Sn.}
        \label{fig:superconductor-field-regimes}
 \end{figure}

The dipole magnets are the main cost drivers for future hadron colliders. 
We note that the infrastructure and expertise needed to make significant progress in this field are the result of 
long-term vision and support from HEP funding agencies. The developments that have occurred in 
high-field dipole magnet performance in the US over the last two decades, shown in Fig. \ref{fig:progress1}, have been 
made possible by the strong support of DOE HEP through the GARD program. The dipole field strengths obtained through these 
programs enables the community to consider collider configurations with energies significantly beyond the LHC, as shown in 
Fig. \ref{fig:superconductor-field-regimes}. Based on the developments 
to date, we can make a general assessment on the current status and potential future of magnet development:
\begin{itemize}
\item The technology basis for dipole magnets generating $\sim15$~T fields (with operating margin) is in place. To bring 
this technology to a state of readiness for application in a future collider will require a focused technology readiness 
effort, similar to that of LARP and of a similar timescale ($\sim 10$ years) and funding level.
\item The technology basis for $20+$~T dipoles is not yet in place, although very promising concepts are being proposed that 
warrant investigation. Strong R\&D support for these efforts, similar to that provided to the magnet research groups over 
the last decade, can be expected to yield results within a $\sim5$ year time frame. If these developments are successful, 
a follow-on phase of technology readiness should be again envisaged, in order to bring this technology to the state needed for 
its implementation in an accelerator project. For these magnets, the conductor developments are critical, and magnet support 
must be matched with appropriate support for conductor R\&D.  
\end{itemize}
The HL-LHC upgrade program requires quadrupoles with peak field in excess of 12~T. It heavily relies on the 
success of the advanced Nb$_3$Sn technology developed by LARP, since the NbTi superconductor used in the present 
LHC magnets is limited to field strengths below 8-9~T~\cite{Ferracin:2010zz, Bottura, todesco}. 
The dipole magnet R\&D is well-leveraged, in the sense that it serves both hadron and muon collider research. We note that 
in addition to field strength, magnet programs need to take into consideration cost-effectiveness and 
scalability in magnet designs. As noted earlier, the superconducting dipoles for a future collider will dominate the 
facility cost. Furthermore, they are one of the highest-risk technical components. Beyond the R\&D phase, designs must be 
industrializable to leverage the strength of the private sector in providing cost-effective fabrication during mass production.

Besides magnets, many other technologies are involved, 
like crab cavities, advanced collimators, SC links, advanced remote handling, etc. The HL-LHC is a medium size 
project and implies deep changes over only about 1.2~km of the LHC ring. Beside its physics goals, the HL-LHC will 
pave the way to a larger project like a higher energy LHC, which will be based on a further development of the same,  
or similar, technologies. In fact, the technological development of high-field magnets and other components discussed below, 
which are key to the HL-LHC program, connect the mid-term upgrade path of the LHC to the long-term developments 
towards hadron colliders of even higher energies. These programs have provided the US with unique know-how at the 
national labs and recognized leadership in magnet R\&D. They also have important overlaps with other HEP research 
areas, such as the muon accelerator program (MAP) and accelerators for the intensity frontier, in the fields of 
magnet and collimation systems. The muon collider design requires dipoles of large field strength (10-20~T) and 
aperture, high-field solenoids and final focus quadrupoles, since the luminosity scales with the available field.  

In summary, it is important to continue a focused integrated program emphasizing engineering readiness of technologies 
suitable for high energy hadron colliders with applications to the muon accelerator program and the intensity frontier.
In this section, we discuss the technological aspects of high field magnets from the conductor development to 
specific designs, the HL-LHC collimation issues and the crab cavity R\&D.

 \subsection{Conductor development for SC magnets}

\begin{figure}
        \centering
        \includegraphics[width=0.55\textwidth]{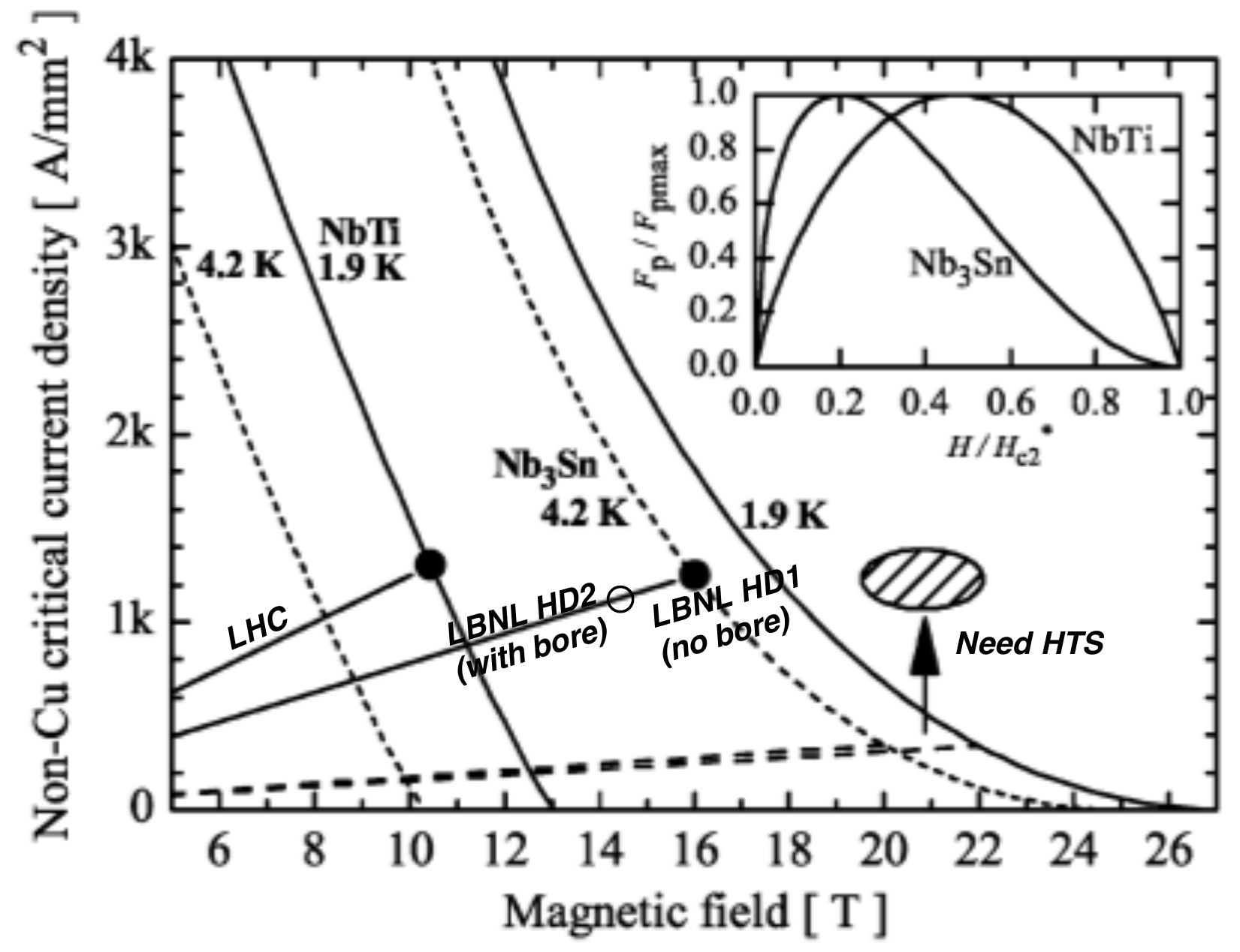}
        \caption{Critical current of NbTi and Nb$_3$Sn at 1.9K and 4.2~K. The inset shows the corresponding normalized pinning force 
         for each conductor. Load-lines for the LHC NbTi dipoles and the LBNL Nb$_3$Sn high-field dipole model magnets HD1 (without bore) 
         and HD2 (with bore) are also given (modified from~\cite{godeke}.}
        \label{fig:Jc-plot}
 \end{figure}

Superconductor properties form the basis of magnet performance and figure~\ref{fig:superconductor-field-regimes} shows .
the role of the superconductor in the energy reach of hadron colliders, across different technologies. The main low-temperature 
superconductors (LTS) used in accelerator magnets are NbTi and  Nb$_3$Sn. 
All SC magnets for colliders to-date have been based on a LTS material, NbTi, which has a critical field $B_{c2}\approx 11$T 
at $1.8$K. NbTi-based accelerator technology has been pushed to its limit in the development of the LHC dipoles, with operating 
fields of $8.3$T at  $1.9$K. 
Since then accelerator magnet research  has focused mostly on another low-temperature superconductor, Nb$_3$Sn, which has a 
much higher critical field, $B_{c2}\approx27$T at $1.8$K (see Figure~\ref{fig:Jc-plot}), and can provide access 
to fields beyond the intrinsic limitation of the NbTi technology \cite{Sabbi:uh}. However, Nb$_3$Sn is a brittle inter-metallic 
compound, which imposes significant constraints on the design, fabrication and implementation of the material in accelerator magnets. 

Improvements in the current density, filament size, stability and production length properties for Nb$_3$Sn has been the focus of the 
GARD-funded Conductor Development Program (CDP), initiated in 1999. CDP provides funds for the support of industrial improvements 
focused on specific target areas. Over the last $\sim15$ years, the industry has made tremendous progress in Nb$_3$Sn. 
The current-density figure of merit $J_c(12T,4.2K)$ has almost tripled to $J_c(12T,4.2K)>3000$A/mm$^2$ (see Figure~\ref{fig:progress1}). 
Ongoing efforts focus on reducing the filament characteristic size, $D_{eff}$, by increasing the number of filaments in the wire, 
and in improving the residual resistivity ratio, RRR, through improvements in the barrier surrounding the filaments. 

A fairly new area of conductor development pertains to high-temperature superconductors (HTS). These materials, primarily Bi$_2$Sr$_2$CaCu$_2$O$_x$ 
(Bi--2212) and \ybco (YBCO),  are under investigation for high-field magnet applications. At low-temperatures (e.g. 1.8-4.5K) they 
exhibit good current densities at high magnetic fields. In fact, $J_c(B)$ is nearly independent of the field value, above $\sim 10$T. 
Over the last few years a collaboration of US national laboratories and university groups has worked to develop Bi-2212 as a viable conductor 
for magnet applications. Funded by the DOE OHEP and first formed as the "Very High Field Superconducting Magnet 
Consortium" (VHFSMC), the group has worked together with industry to develop a baseline round wire. It was able to demonstrate that the wire could 
be cabled into the standard Rutherford cable configuration for a current-scalable conductor. It also proceeded to design, fabricate and test the 
first solenoid and racetrack coils. This effort demonstrated the basic viability of the conductor. However, it also made clear that the effective 
("engineering") current density ($J_E$) of the wire was not suitable for accelerator magnet application. 
Therefore, a smaller, conductor-focused group, the "Bi-2212 Strand and Cable Collaboration" (BSCCo), was formed to focus on the $J_E$ of Bi-2212. 
This effort was successful in demonstrating that the $J_E$ could be tripled, by modifying the processing to include high-pressure during the heat 
treatment, thereby dramatically reducing porosity in the final conductor~\cite{Larbalestier:2013lwa,shen}.
This development is now the focus of renewed interest by the magnet community, with active programs in place to demonstrate dipole and solenoid 
inserts suitable for accelerator applications.

\subsection{High field magnets for HL-LHC}

\begin{figure}[h!]
\begin{center}
\includegraphics[width=0.65\textwidth,clip]{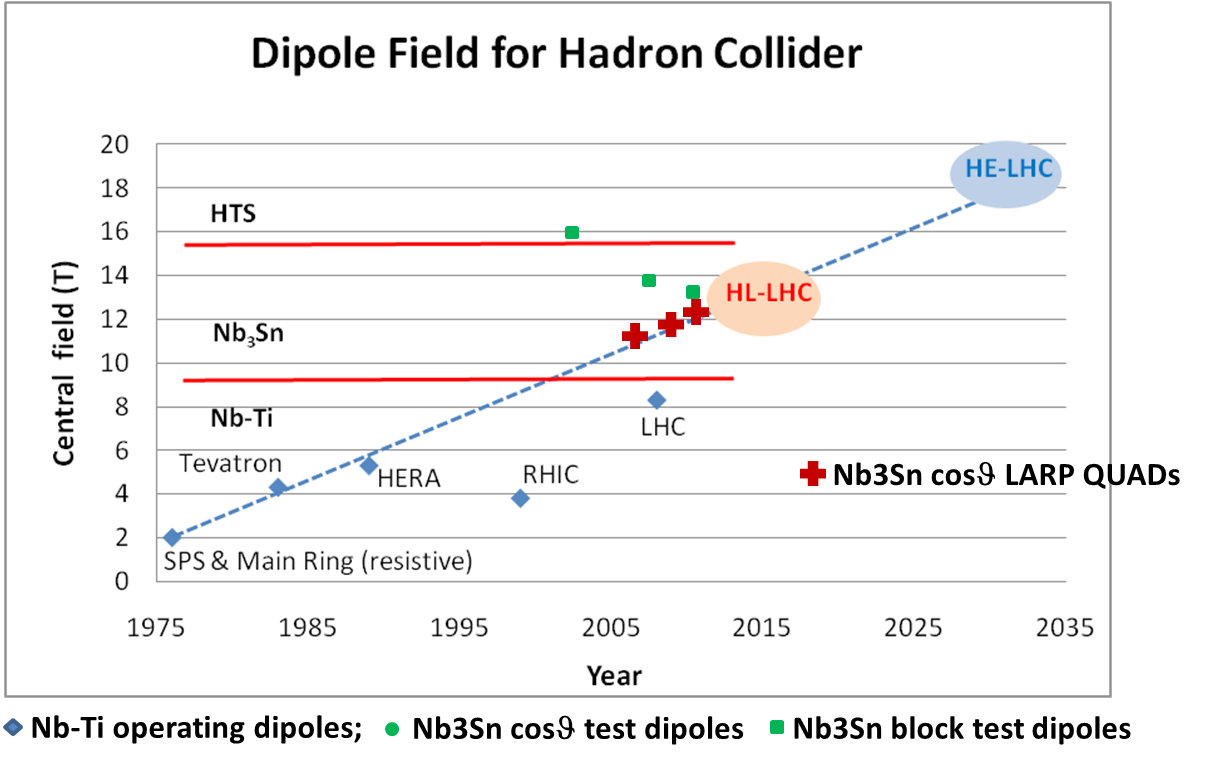}
\end{center}
\caption{Progress of accelerator magnets for hadron colliders. Below 9~T is the realm of Nb-Ti, nominal operating condition of Nb-Ti 
magnets used in the present machines arwe given, Above 9 Tesla, Nb$_3$Sn is required. Results refer to prototype magnets and 
the record achieved field, rather than the operating value, is given. Early Nb$_3$Sn dipoles have 50~mm bore while the HD 
series features an aperture of 40~mm. LARP magnets are quadrupoles, with 90 (LQ) or 120 (HQ) mm aperture. For comparison, 
HL-LHC requires a quadrupole aperture of 150~mm.}
\label{fig:4} 
\end{figure}

The progress over time of accelerator magnets for hadron colliders is summarised Fig.~\ref{fig:4} from the resistive magnet era with 
the jump in performance required by HL-LHC. The graph summarizes also R\&D results and requirements towards a high energy machine, 
that will be discussed below.

However, field strength alone is not sufficient~\cite{rossi}. 
Radiation escaping from the collision point through the beam pipe has two main effects: i) heat deposition 
that may limit the performance of the SC magnets by increasing the conductor temperature; 
ii) radiation damage, especially to insulation but also to metallic components. 
Radiation damage for magnets exposed to the radiation near the experiments is directly 
proportional to the integrated luminosity and might occur around an integrated luminosity 
of 300 fb$^{-1}$, which is probably a conservative  estimate. Heat deposition may limit 
the peak luminosity at about 1.7 to 2.5 $\times$ 10$^{34}$cm$^{-2}$s$^{-1}$. In both cases 
this means that sometimes after 2020 the low-$\beta$ quadrupole triplets will need to be replaced. 
This is a unique chance to install magnets of new generation, capable of higher performance to make the full 
system more robust against radiation and other factors, which would otherwise reduce the availability of the machine.
In addition to the replacement of the quadrupole, the whole Interaction Region (IR) 
zone needs to be redesigned with larger aperture recombination/separation 
dipole magnets (D1/D2 pair), new distribution feedboxes (DFBX), cryo-distribution electrical feed-box of 
the low-$\beta$ triplet and improved access to various equipment for maintenance. The cryogenic system 
will also need to be improved and the power supplies and distributed feed-back (DFB) most critical in terms 
of radiation exposure should be placed out of the tunnel. Power supplies and DFB would be relocated on the surface, 
by means of powerful 150~kA SC links, to reduce the time of intervention and make maintenance easier. 
Many other systems, such as SC magnet protection, interlock, etc., will also need to be upgraded to improve 
the machine availability, essential to achieve high integrated luminosities, as mentioned above.

\begin{figure}[ht!]
        \centering
        \begin{tabular}{cc}
        \includegraphics[width=0.275\textwidth]{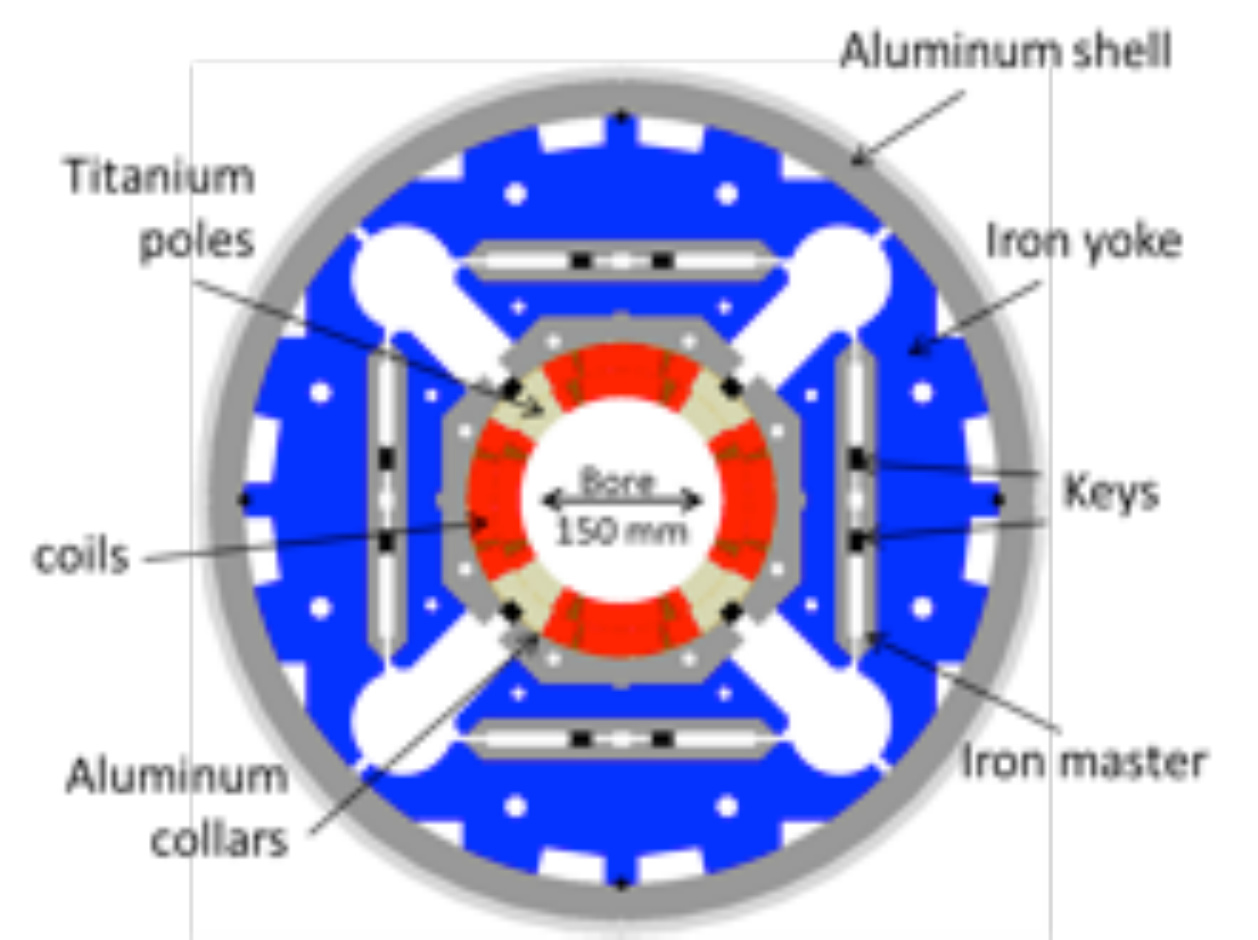} &  
	\includegraphics[width=0.625\textwidth]{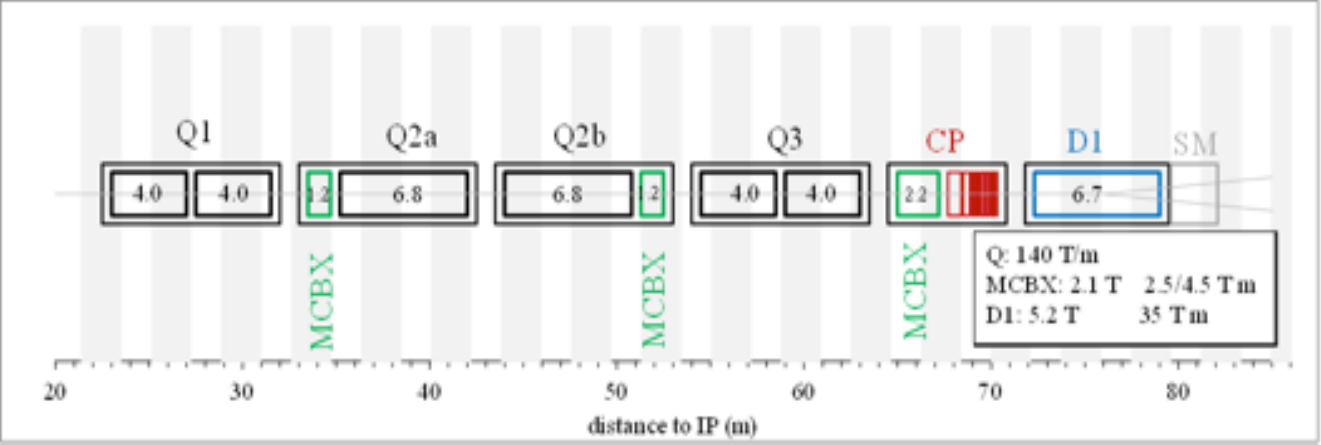}
        \end{tabular}
        \caption{Left: cross section of the 150~mm aperture quadrupole of the inner triplet, under development by CERN and 
        the US LARP program. 
	Right: layout of the HL-LHC IR (inner triplet, correctors and D1).}
        \label{fig:triplet}
 \end{figure}

The magnets for use in accelerators need a current density between of $J_c \simeq$ 2500-3000 A/mm$^2$ , approximately 3 times 
larger than those for the ITER fusion program. This has been possible thanks to the long-term Conductor Development 
Program. Once the conductor becomes available, the magnet field progresses steadily as shown by the progress of 
the maximum field in short Nb$_3$Sn dipole or quadrupole magnets (what we call models, typically 1 m long or less). 
Nb$_3$Sn record values of the field strength are obtained after many quenches and in conditions far from 
operation in an accelerator. Still, the performance required at the HL-LHC is now within reach. 
The LARP quadrupoles have reached the 12~T field level on the conductor with an aperture of 90-120~mm, much larger than that 
of the present LHC quadrupoles (56-70~mm) and already very close to the 150~mm aperture which is the target value for the HL-LHC.
In Fig.~\ref{fig:triplet} the new triplet scheme for the HL-LHC upgrade is shown. 
Special tungsten shielding will be placed inside inner bore to limit the radiation deposition to the same 
level of the nominal LHC, about 30~MGy, despite the ten time higher integrated luminosity.
The US material contribution to the HL-LHC project may consist in the production of half of the new quadrupoles 
(Q1 and Q3 for the four sides of the two interaction points). 

\subsection{High field magnet R\&D for pp colliders beyond the LHC}

\begin{figure}[h!]
        \centering
        \includegraphics[width=0.3\textwidth]{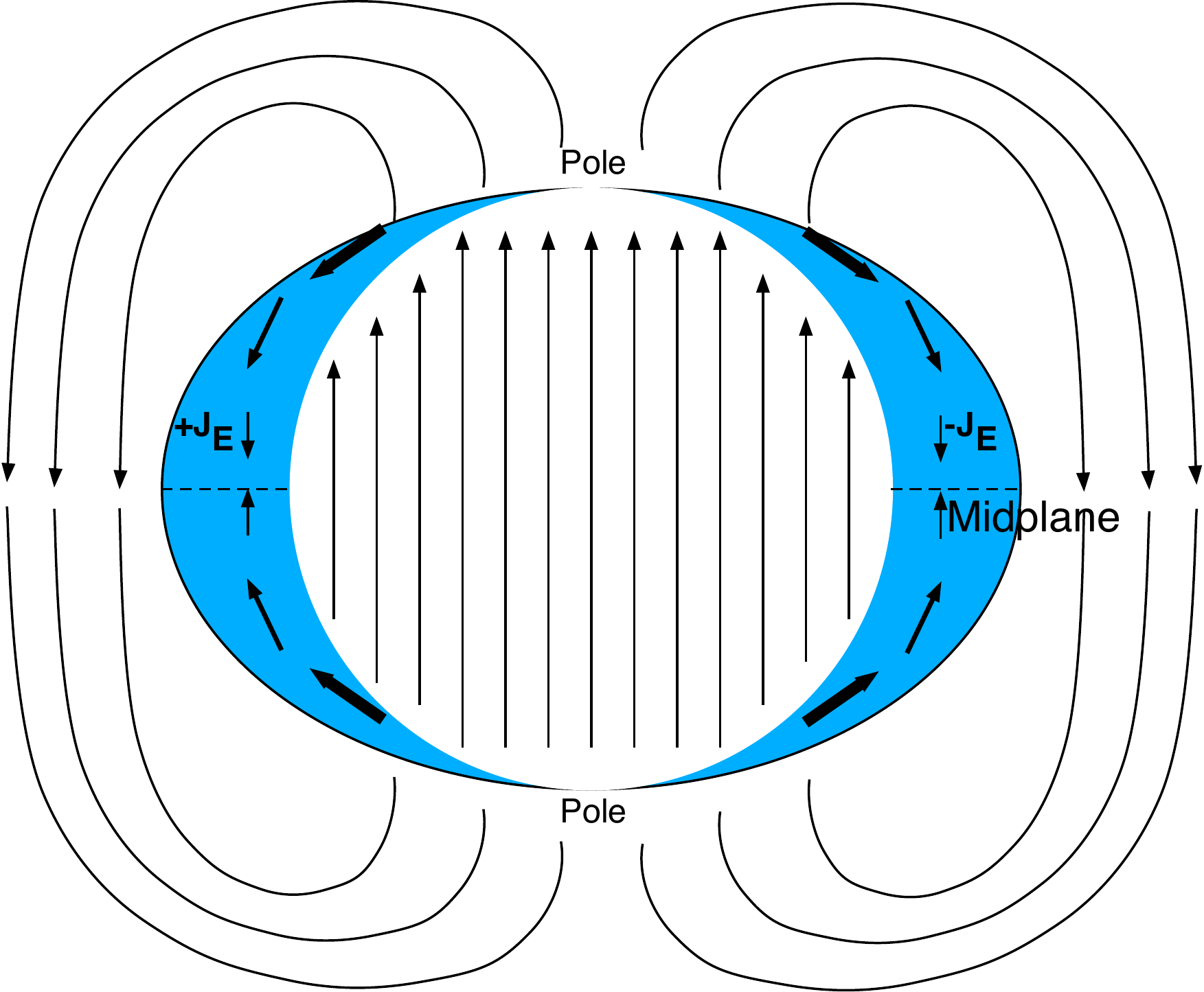}
        \caption{Schematic of a $Cos(\theta)$ current density distribution and azimuthal forces. Note that the forces accumulate on the azimuth, resulting in peak 
	stress on the mid plane of a $Cos(\theta)$ dipole.}
        \label{fig:cos-t-azimuth}
 \end{figure}

In order to understand the issues driving high-field dipole development, it is helpful to consider here the fundamental design limitations. 
The concept of a $Cos(\theta)$ dipole is shown in Fig. \ref{fig:cos-t-azimuth}.  

\begin{figure}[ht!]
        \centering
        \includegraphics[width=0.85\textwidth]{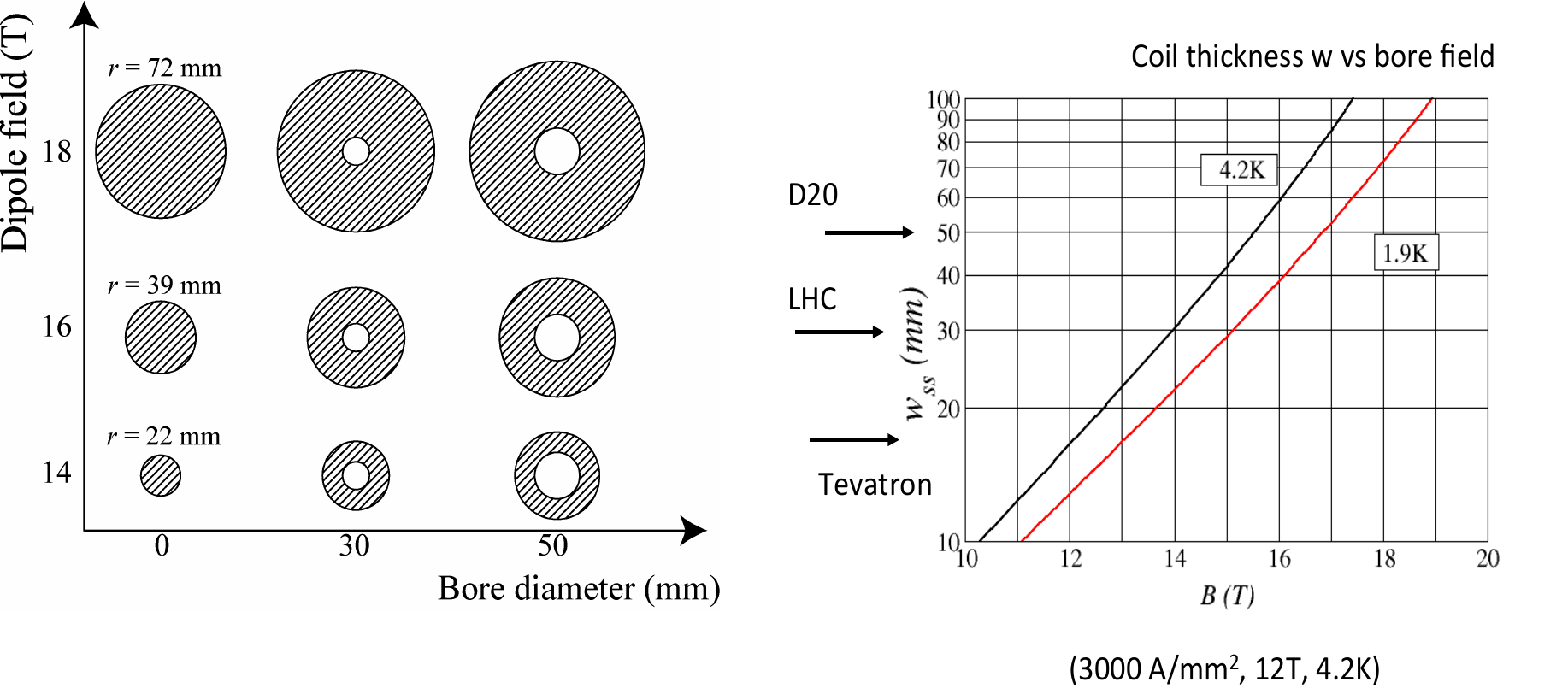}
        \caption{The magnetic field produced by a $Cos(\theta)$ current density distribution is a function of coil width and current density, but not of the coil radius.}
        \label{fig:coil-thickness}
 \end{figure}

The field produced by a perfect $Cos(\theta)$ current density distribution is 
a function of the current density $J_E$ and coil width $w$, and independent of the radius $r$ as shown in  Fig. \ref{fig:coil-thickness}. 
  \begin{figure}[h!]
        \centering
  \begin{tabular}{cc}
        \includegraphics[width=0.45\textwidth]{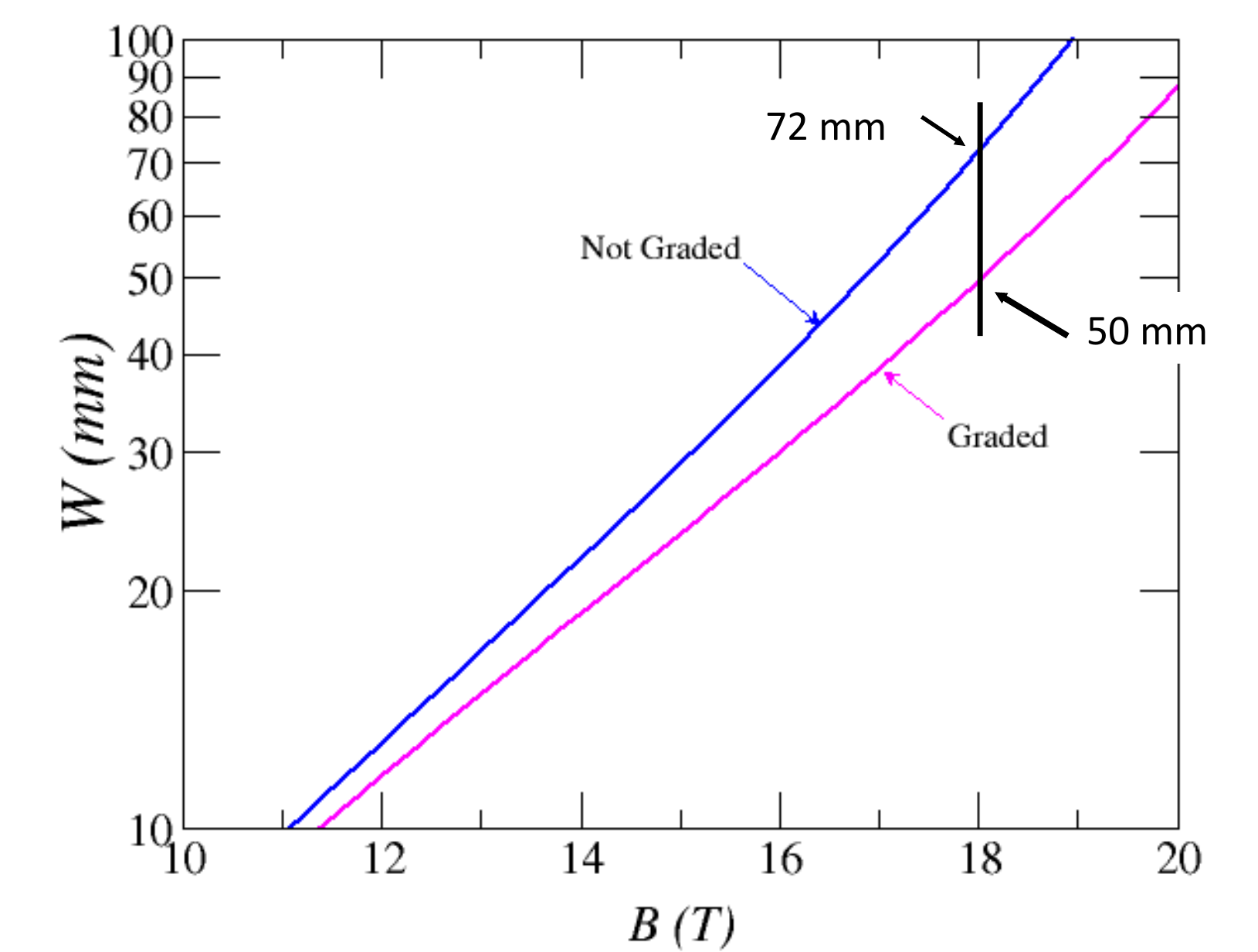} &
        \includegraphics[width=0.45\textwidth]{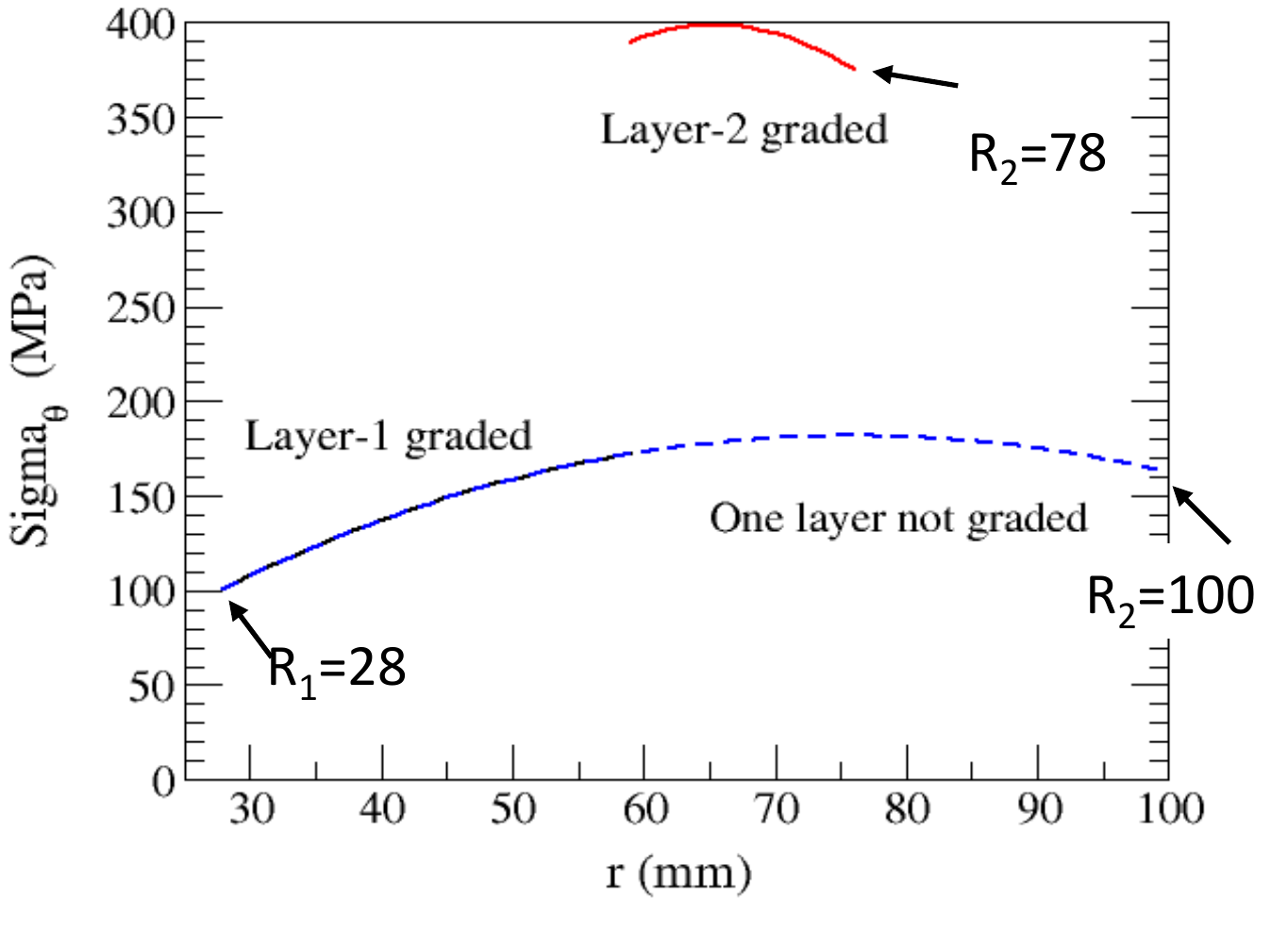} \\
  \end{tabular}
        \caption{(Left) Improvement in field-generating efficiency of  $Cos(\theta)$ current density distribution using "grading". 
                 (Right) Enhancement of $\sigma_{\theta m}$ associated with grading of a  $Cos(\theta)$ magnet. Values correspond 
		 to a $\sim18$T Nb$_3$Sn scenario.}
        \label{fig:grading}
 \end{figure}
 
Hence the volume of conductor ($V$) needed to produce a field $B_0$ scales like $V\propto r$. This scaling can be improved by "grading" the conductor, i.e. leveraging the $J_c(B)$ behavior of superconductors to  tailor the current density of the wire as a function of radial position, resulting in less conductor volume for a given field. 
The value of grading with Nb$_3$Sn is shown in Fig. \ref{fig:grading}. The Lorentz force $F_{\theta}\sim J_E B_r$ accumulates along the azimuth, resulting in maximum 
stress $\sigma_{\theta m}$ on the midplane (see Fig. \ref{fig:cos-t-azimuth}). Experience with Nb$_3$Sn magnets has lead to well-established limitations $\sigma_{\theta m}<200$MPa on the compressive mid plane stress. Beyond this value the conductor degrades and magnet performance becomes unreliable. Due to the accumulation, stress on a 
$Cos(\theta)$ coil of radius $r$ and thickness $w$ scales as 
 \begin{equation}
 \sigma_{\theta m}\propto r J_E/w.
 \label{eq:stress-scaling}
 \end{equation}
Implementing grading in a high field $Cos(\theta)$ dipole results in dramatic increase in $\sigma_{\theta m}$, since $w$ decreases and $J_E$ increases with $r$ when 
grading is incorporated. Fig. \ref{fig:grading} provides an example for a Nb$_3$Sn $\sim 18$T coil layout. It is important to note that the graded scenario far exceeds 
the stress limitation for the material.  
 
\begin{figure}[h!]
        \centering
        \includegraphics[width=0.5\textwidth]{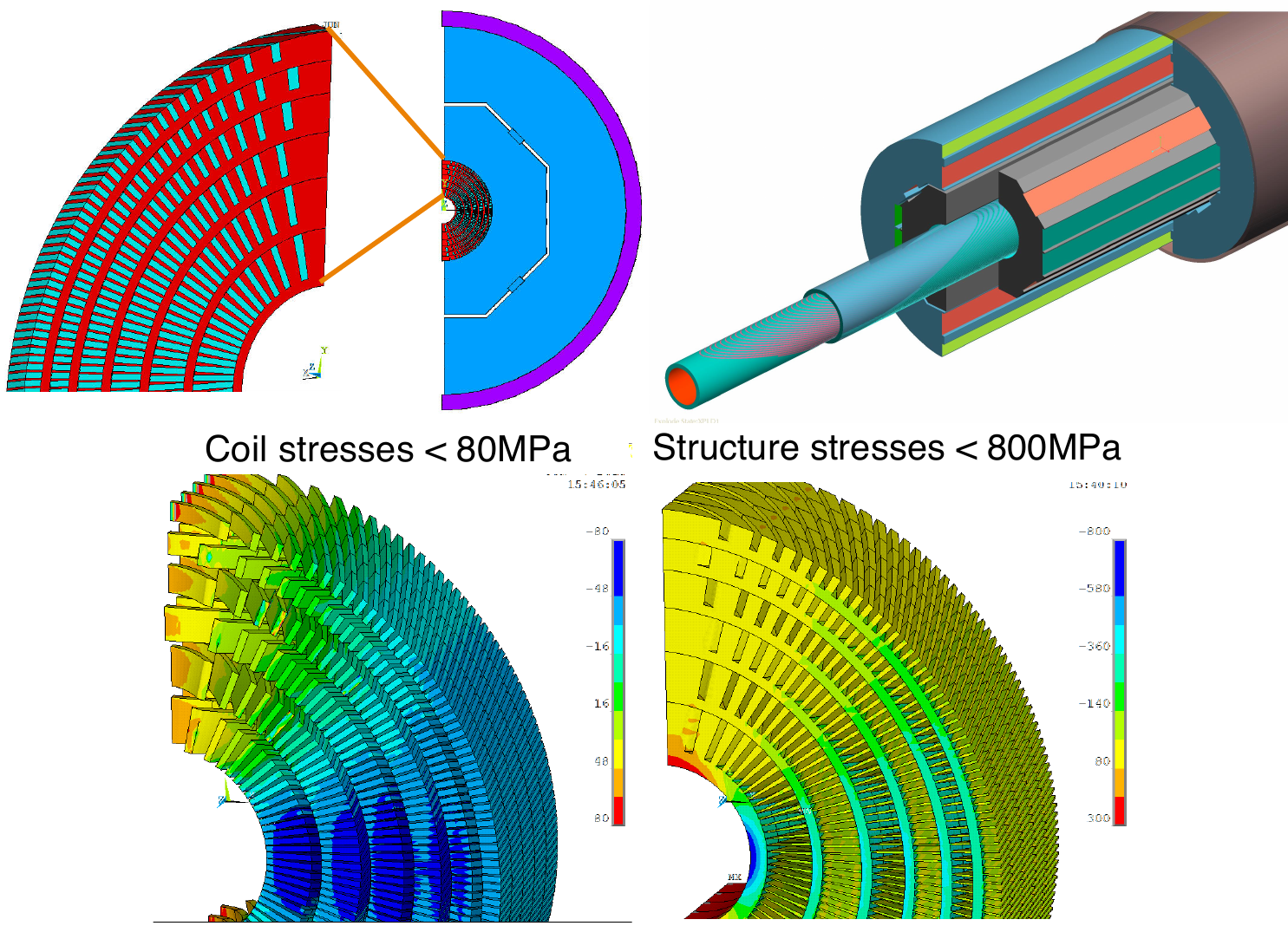}
        \caption{Concept of a Canted $Cos(\theta)$ magnet yielding 18~T. The design, composed of 6 layers, uses graded Nb$_3$Sn Rutherford cable. 
	Azimuthal forces acting on each cable are captured by the ribs. Radial forces are supported via prestress from the external structure. 
	Peak stresses on the conductor are significantly lower than on existing high-field dipoles with a reduction of an order of magnitude compared to an 
	equivalent $Cos(\theta)$ design.}
        \label{fig:CCT-v2}
 \end{figure}

 Stress is a dominant limitation for high-field dipole designs. From Eq. \ref{eq:stress-scaling} it is evident that $\sigma_{\theta m}$ can be reduced to acceptable values by reducing the current density and making the coil cross-section larger. This comes at the expense of significant more conductor to be used. Furthermore, the actual stress 
state in the $Cos(\theta)$ design is even a more complex phenomenon than it was described above, due to significant radial forces acting on the coils, predominantly in the 
midplane region. 
 The importance of stress accumulation in dipole magnets has been recognized for some time. Alternative magnet designs, using "block" coils in various configurations, has been the focus of much effort over the last $\sim15$ years. A concept of stress management was proposed by the  Texas A\&M group \cite{Elliott:1997us}. The idea is to intercept stress before it can accumulate to unacceptable values. An alternative approach, using a "canted $Cos(\theta)$" design concept (see Figure~\ref{fig:CCT-v2}), is presently under investigation at LBNL \cite{Meyer:1970}.
 
\subsubsection{Magnet protection}

Another important limitation of high-field dipoles and quadrupoles concerns the stored magnetic energy, $E_m\propto \pi r^2 B^2_0 L$, 
where $L$ is the magnet length. The use of high current density superconductors, results in significant heating of the magnet in the case of a quench, 
i.e.\ the sudden irreversible loss of superconductivity at some point in the conductor. Highly specialized magnet protection circuitry is now employed 
in high-field accelerator magnets, typically incorporating the following features:
\begin{itemize}
\item fast detection of a quench onsite, typically within a few milliseconds  of quench initiation - the detection must discriminate between real quench signals and false triggers emanating from flux jumps that can randomly occur;
\item firing of heaters distributed in the magnet to force a large fraction of the superconductor into the normal (non-superconducting) state and to force 
the drive current to flow in a parallel bus bar line; 
\item activation of an external dump resistor to extract a significant fraction of the power away from the magnet and to protect the parallel bus bar line. 
\end{itemize}
All of these events must happen on a timescale of $\cal{O}$(10~ms) in order to prevent areas of the magnet from overheating. As the current densities in superconductors 
improve, and higher-field magnets with larger bore are developed, magnet protection becomes an increasingly  significant technical challenge.

\subsubsection{Field quality}

Dipole magnets for a future collider will have strict requirements on multipole content and hysteresis. These issues are impacted by the choice and characteristics of the superconductor and by the magnet design. SC wires are composed of a large number of filaments that are then twisted to minimize coupling losses, i.e. resistive Joule heating generated by $dB/dt$ during magnet ramping. The filaments themselves can support persistent currents that impact field quality and are hysteric in nature. Furthermore, the filaments can experience sudden flux-penetration ("flux-jumps") as fields vary and the shielding nature of the superconductor is overwhelmed by the local Lorentz force acting on the pinned vortices. Although in most cases the superconductor can recover from such events, the resulting flux dynamics cause small field fluctuations that must be minimized.

These issues are addressed by minimizing the  effective diameter of the filaments. This is an area of active development by industry, with support from the GARD CDP in the US. We note that, at present, the HTS materials, Bi-2212 and YBCO, both suffer from large effective filament sizes. Significant advances will have to occur in the material technology for the conductors to be viable candidates for accelerator dipoles. Other magnet systems, such as the interaction region (IR) quadrupoles and the solenoids for a Muon collider, are much less sensitive to field quality concerns and may be the early adopters of HTS materials. It should be emphasized that the use of high $T_c$ conductors gives an important temperature margin to these systems and therefore needs to be a priority in the US HEP R\&D program.
  \begin{figure}[h!]
        \centering
        \includegraphics[width=0.5\textwidth]{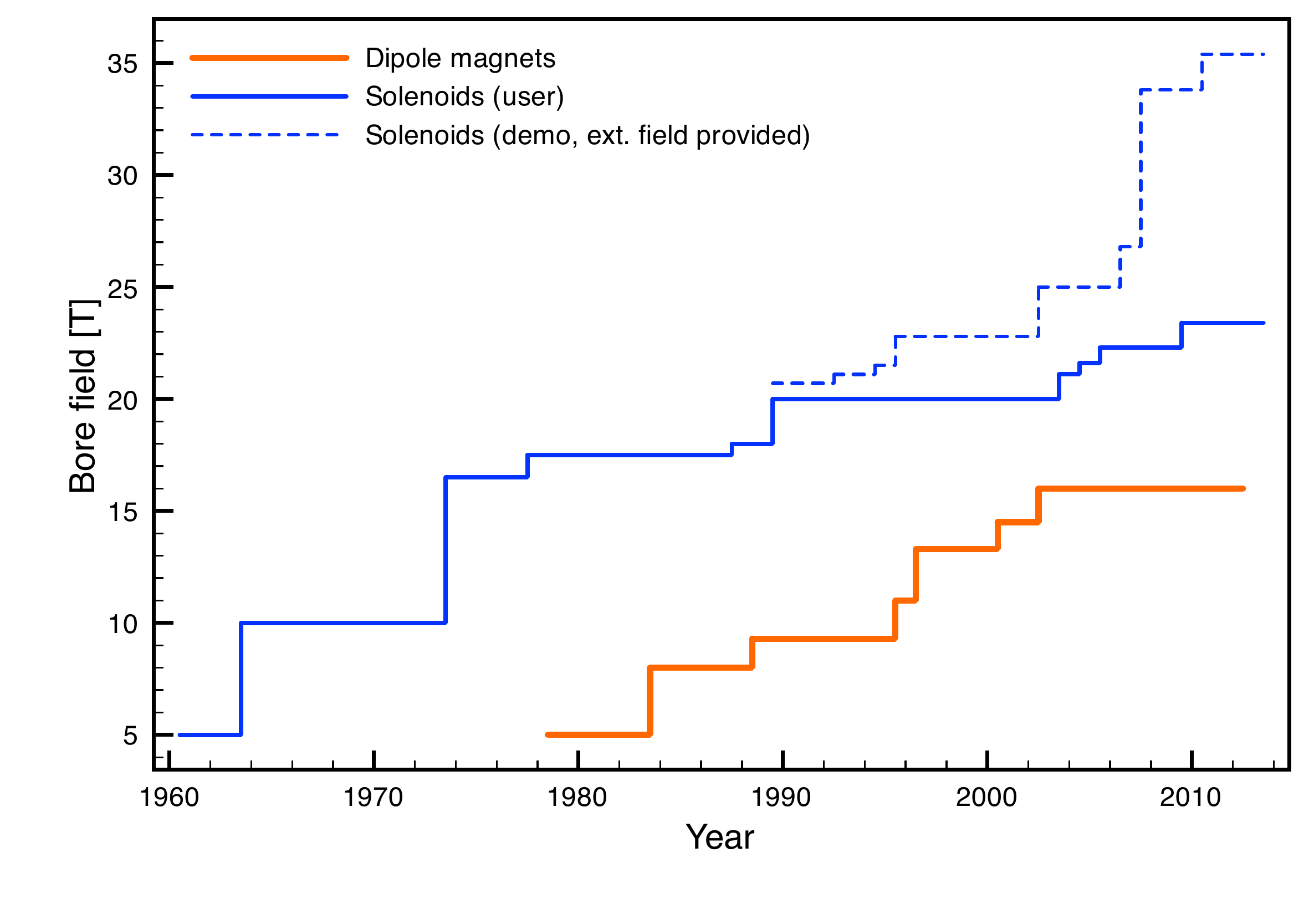}
        \caption{Progress in SC dipole magnet performance vs solenoids over the last few decades. Note the step increases, typically associated with significant material developments and/or new design paradigms. The dotted line represents prototype solenoids, where SC solenoids are tested in a background field provided by resistive magnets.}
        \label{fig:magnet-evolution}
 \end{figure}

 In terms of magnet design, field quality is dictated by the conductor layout in the coil cross section, a fairly well understood area. Open issues 
include the understanding of: 
 \begin{itemize}
 \item the fabrication and assembly tolerances and their impact on field quality;
 \item the influence of thermal contraction and deflections associated with magnet energization, and their impact on field quality.
 \end{itemize}

\subsection{Collimation}

The LHC collimation system is presently operating well but, in order to cope with higher beam energy densities as well as 
to lower impedance, will have to be renovated. Gains in triplet aperture and performance must be matched by 
an adequate consolidation or modification of the collimation system. In addition, a collimation system in the 
dispersion suppressor (DS) needs to be added to avoid the leakage of off-momentum particle, into the first and 
second main SC dipoles. This has been already identified as a possible limitation for the LHC performance. The most 
promising concept proposed so far requires to replace one of the LHC main dipoles with a dipole of equal bending 
strength (121 T-m), shorter length (11~m) and higher field (11~T) compared to the dipoles (8.3~T and 14.2~m) currently 
installed in the LHC. The resulting gain in space is sufficient for placing special collimators to intercept the 
off-momentum particles. The new 11~T dipole, jointly developed by CERN and Fermilab, will be actually realized with 
two cold masses of 5.5~m length, and should become the first magnet breaking through the 10~T field strength frontier 
to be installed in a particle accelerator.

\subsection{Crab cavities}

\begin{figure}[h!]
        \centering
        \includegraphics[width=0.65\textwidth]{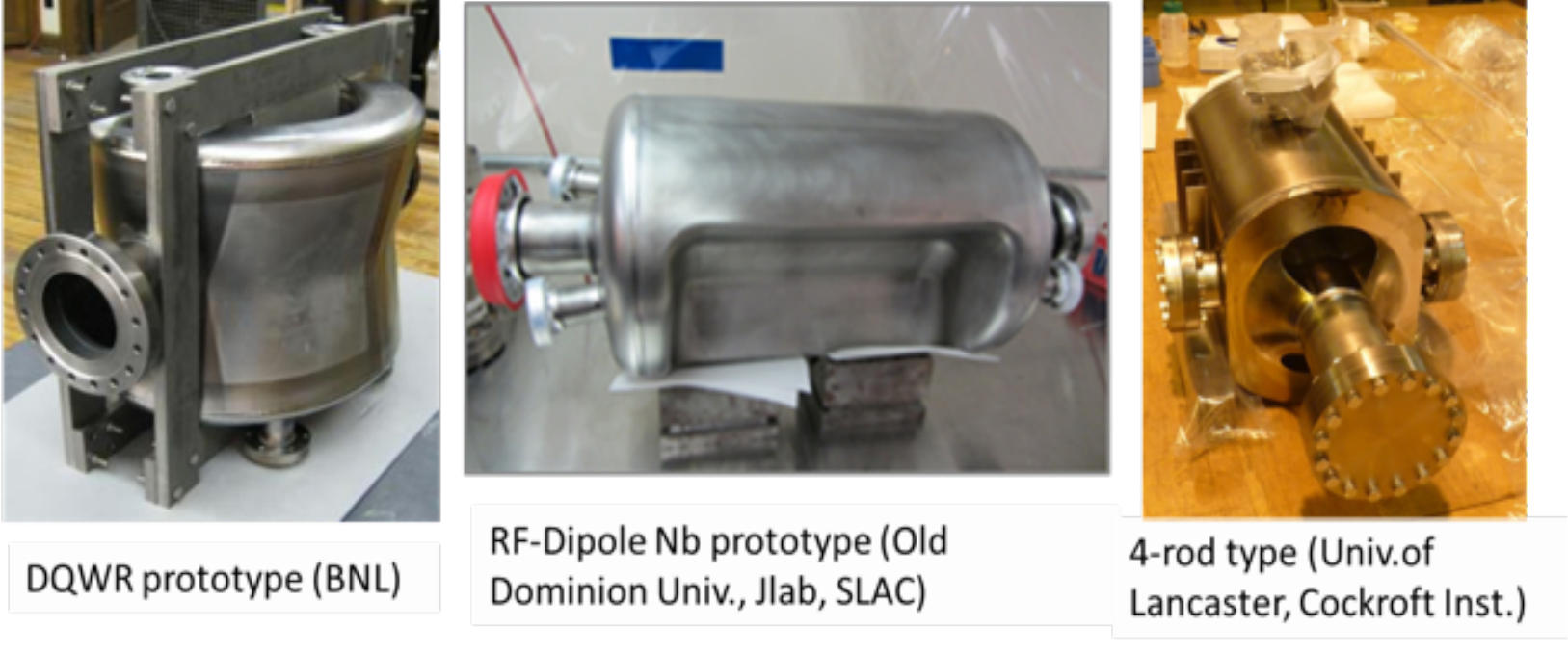}
        \caption{The three types of compact crab cavities under development for HL-LHC.}
        \label{fig:crabs}
 \end{figure}

In order to preserve the luminosity for beams colliding at a large crossing angle, the beams should be rotated before collision  
using RF cavities operating in a magnetic field, known as crab cavities. Crab cavities have been successfully installed at 
the KEKB $b$ factory~\cite{Abe:2007bj}. The crab cavities foreseen for the HL-LHC upgrade are not particularly demanding in terms of 
the required voltage. However, they go beyond the state-of-the art because the transverse cavity dimension is limited by 
the small distance between the two LHC beams (194~mm), which is smaller than the $\lambda$/4 value of the 400~MHz wave. 
This practically excludes the elliptical cavity geometry adopted at KEKB. The need for a small beam separation calls for an 
unconventional, very compact design for the HL-LHC crab cavities. Different solutions have been proposed. After detailed studies 
and R\&D steps, three realistic design options have emerged, which are presently under developement through prototypes, shown 
in Figure~\ref{fig:crabs} 

\begin{figure}[h!]
        \centering
        \includegraphics[width=0.60\textwidth]{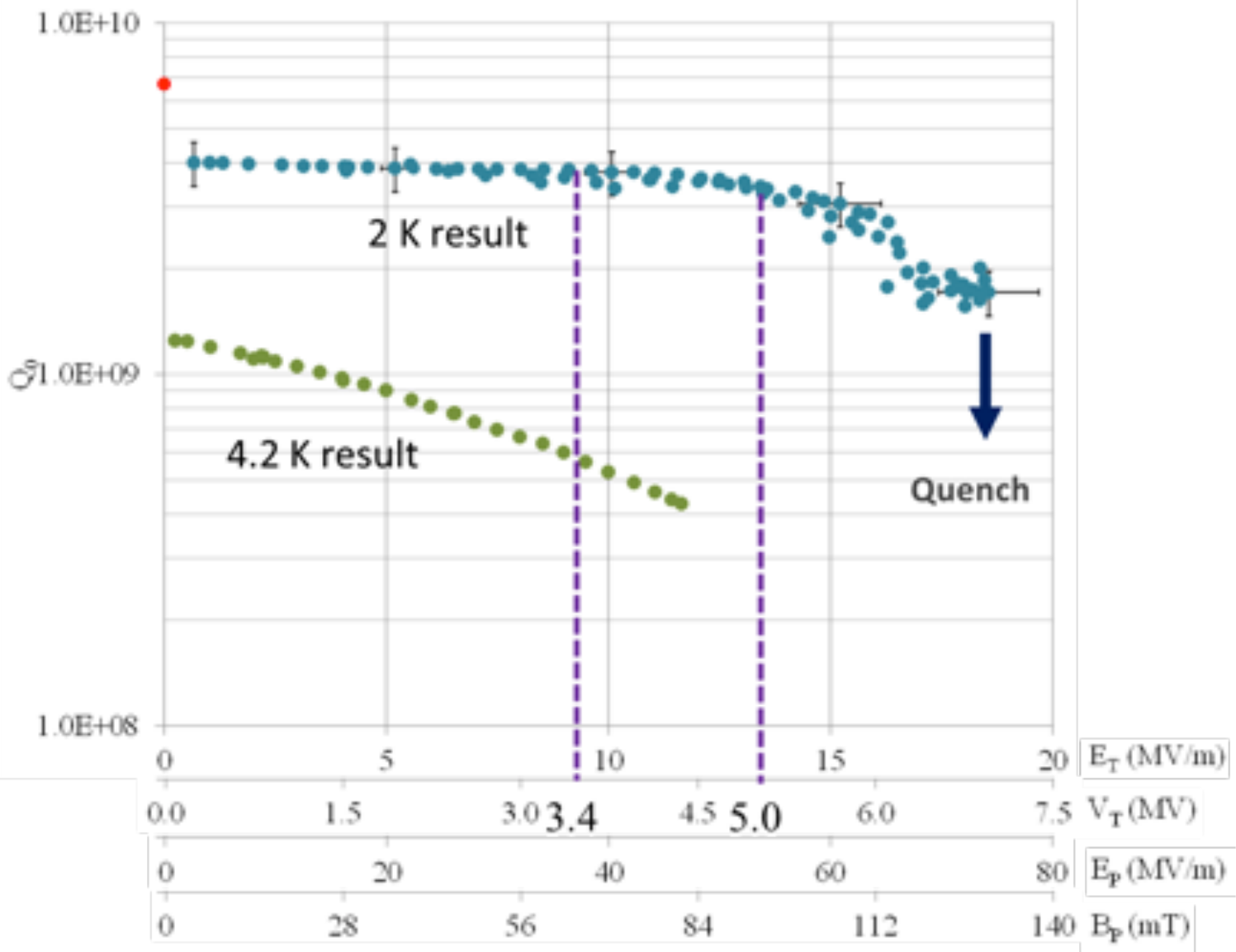}
        \caption{Results of the first full test of a crab cavity: the RF dipole Nb prototype (ODU-SLAC, LARP). The vertical 
lines show the target voltage ($3.4$ MV) and the actual usable voltage ($5$ MV). Beside the transverse gradient and 
voltage, on the horizontal axis, the peak electric field and magnetic field are indicated. The test has been carried 
out at Jefferson Lab.}
        \label{fig:crab-test}
 \end{figure}

The first full test of a cavity prototype has just been completed, for the RF-dipole type. It was operated well above 
the target value of 3.4~MV transverse voltage VT (see Figure~\ref{fig:crab-test}), 
quenching at 7~MV and showing an operational gradient of 5~MV. This result is very encouraging, although several issues remain to be 
addressed. These include their effect on the proton beam in term of noise-induced beam emittance growth and the understanding of the 
possible failure modes, which must be carefully  studied in order to allow the safe operation of the machine~\cite{Calaga:2011zc}. 
The cavity integration in a very compact cryostat must also be addressed. 

Therefore, the possibility of adopting crab cavities at the HL-LHC is not yet full demostrated. Should the use 
of crab cavities not be feasible, the loss in luminosity could be minimized by modifying the beam parameters to 
collide flat beams at the smallest possible crossing angle and pushing the compensation for the long-range beam-beam 
interactions by installing compensating conducting wires, where the current counteracts in part the beam-beam effects.

\section{Higher energy colliders}
\label{sec:vlhc}



Given the progress in magnet technology and the maturity that Nb$_3$Sn is reaching, through the development for 
the HL-LHC, it is legitimate to forecast that Nb$_3$Sn magnets can attain their limit of 15.5-16~T in field 
strength, with a 15-20\% margin in operative condition, within the next decade. This opens the path towards a 
collider with energy significantly larger than that of the LHC and brings an exceptional potential for further 
probing the energy frontier. 

\subsection{HE-LHC}

\begin{figure}[h!]
    \centering
    \includegraphics[width=0.95\textwidth]{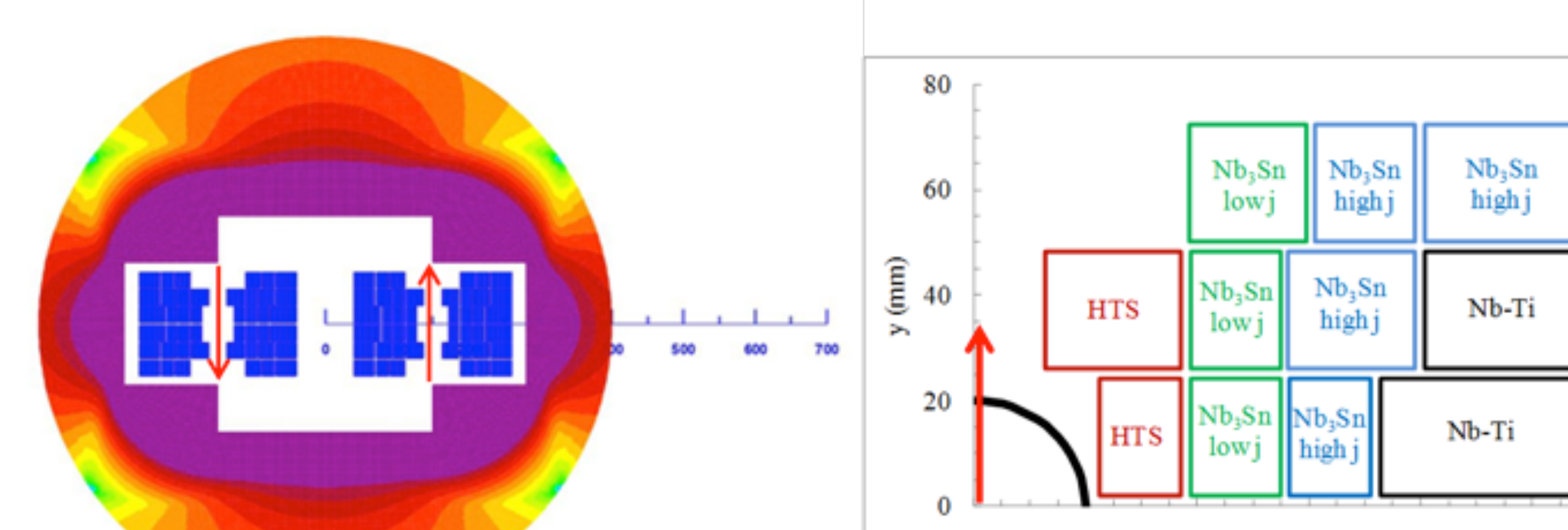}
    \caption{Concept of a Hybrid block-dipole magnet yielding 20~T (from \cite{dipole}). The design uses YBCO coils in the high-field region, 
    Nb$_3$Sn in the "middle" field region, and NbTi in the low-field region. The design, which is only 2-D at this stage, meets the basic requirement 
    of field quality and mechanical stress. It still needs a thorough evaluation in many aspects, including construction technique, peak stresses 
    and magnet protection.}
    \label{fig:CERN-20T}
\end{figure}

The first option being considered is a machine of higher energy housed in the present LHC tunnel (HE-LHC). The achievable 
center-of-mass energy depends on the dipole field strength. An energy of 26~TeV is within reach using Nb$_3$Sn technology 
magnet, which offers a big advantage in reducing complexity and cost of the magnet system. The center-of-mass energy would 
become $\sim$33~TeV,  if 20~T dipole magnets were available, which would require the use of more futuristic HTS conductors.
In this case, the limited space available in the LHC tunnel represents another important challenge to overcome. The 
inter-beam distance must be increased to 300~mm in order to allow for thicker superconducting coils in the main 
dipoles and quadrupoles in order to generate the required 20~T field. 
A 2D design of the magnet featuring an aperture of 40~mm and an operative field of 20~T with realistic overall 
current density in all coils block ($J_E$ = 400~A/mm$^2$) has been initiated. 
The inner block of HTS adds about 4-5~T, and it is necessary for the 20~T target (see Figure~\ref{fig:CERN-20T}). 

Anti-coils are needed to reduce the stray field and the collider field quality is not yet proved. In addition, the magnet 
powering and protection will certainly require significant development. A vigorous R\&D program is therefore  
needed to demonstrate the viability of HTS-based cables and the subsequent magnet engineering design. 

The beam parameters for the HE-LHC are not too different from those for the LHC or HL-LHC. In this 
respect the machine design looks feasible~\cite{helhc,bruning}. However, there are two areas which have been identified as 
critical, in addition to the magnets. First, the injection scheme relies on kickers more powerful than those used at 
LHC, which are already at the limit of current technology. The extraction system is also critical, though to 
a lesser extent. An important program of R\&D in fast-pulsed, high-strength kickers is a necessary complement 
to the dipole magnet program. Then, the beam pipe and beam screen have to absorb a synchrotron radiation which 
is more than 20 times higher compared to the LHC. Synchrotron radiation is very beneficial for beam stabilization 
and would make the HE-LHC to be the first hadron machine dominated by synchrotron radiation dumping. However, the 
power dissipated in SR must be removed at cryogenic temperature. In LHC it is removed at 5-10~K. 
A similar solution with a beam screen at 10~K will be a heavy burden for the cryogenics. The additional 12~kW of power 
needed for each of the eight sectors would require to roughly doubling the present 8$\times$18~kW cryoplants. In addition, 
the beam screen refrigeration would be complicated by the need to increase the local heat removal by a factor of 20 by 
increasing the pressure drop and the conductance of the cooling pipe. Although this solution should be possible, better 
options seem to be available.

The first option would be to remove the SR heat at higher temperature. Vacuum stability indicates two possible windows for beam 
screen operation: 40-60~K (inlet-outlet temperature) and 85-100~K. The first window of operation would maintain the cryogenic power 
at 4.2~K to be equal to that of the LHC; the second window would make refrigeration easier than in the baseline LHC. However, the 
first option is preferable because in the second option the heat leakage from a higher temperature beam screen onto the 1.8~K cold 
mass would be more than double that of the first option. Moreover the electrical resistivity increases by a factor of 5 above the 
LHC value for the first option and a factor of 22 for the higher temperature window.

The consequences of the higher resistivity on beam stability are not dramatic, because both the transverse and the longitudinal beam 
impedance increase with the square root of the wall resistivity. The resistivity of copper at 50~K and 20~T is just a factor 2.5 more 
than that at 20~K, 8.3~T, resulting in an impedance increase of $\simeq$60\%, which is a manageable factor. A thicker copper coating 
than the 75~$\mu$m used in LHC will partially compensate this increased resistivity, compensation necessary to cut down power losses 
due to image currents.

Although a copper coating appears to be a viable solution for HE-LHC, the use of YBCO coating ($T_c$ = 85~K) on the inner surface of the 
beam screen (if practical) could virtually null the resistance, eliminating the problem. The HTS coating will even make possible working 
at 100~K (if the thermal contact to the 1.9~K cold mass can be made tiny enough).  This HTS coating will certainly   be more expensive and 
complex that the copper-stainless steel co-lamination of the LHC beam screen. However, given its potential benefits, it should be carefully 
investigated as part of this R\&D program.

\subsection{Proton Colliders beyond LHC}

Studies for a very large proton collider able to deliver center-of-mass energies of order of 100~TeV (and more) have been 
already conducted since the mid-1980s and subsequently updated, in a detailed multi-laboratory study led by Fermilab~\cite{vlhc}
An update based on higher field magnets has been prepared and submitted as a white paper to this Community Summer Study~\cite{Bhat:2013epd}.
A study to investigate the feasibility of a 80-100~km ring for an next-generation hadron collider (VHE-LHC) is now starting at 
CERN, with the aim to produce a complete machine conceptual design by 2018-20. 

The construction of a new tunnel relieves the pressure on the achievable dipole field strength, 
since this can be traded for the tunnel circumference. The target collision energy is 100~TeV for 20~T dipoles in 
an 80~km tunnel.  However, an 100~km tunnel would provide the same collision energy of 100~TeV with reduced field of 
16~T, reachable with Nb$_3$Sn technology which a much more mature and less expensive conductor than HTS. 
The CERN civil engineer team has studied a 5~m diameter tunnel, compared to the 3.8~m of the present LHC tunnel thus allowing 
for larger cryo-modules~\cite{osborne}.
Cost and magnet technology set aside, the main issue for the VHE-LHC collider is the removal of synchrotron radiation. 
In a 80~km,  100~TeV proton collider, the dumped SR power will jump from the 3.6~kW/ring of the LHC and 82~kW/ring of a 
HE-LHC to $\sim$2~MW/ring. Dealing with this 500 times increase of SR compared to the present LHC value will be a major issue. 
While the beam stabilization will get a tremendous benefit, making the machine much easier from the point of view of the 
beam dynamics, we do not know if a beam screen cooled at 100~K (to limit thermodynamic load) would be capable to withstand 
the resulting heat deposition. Furthermore, the critical energy of the emitted photon is in the soft X-ray range and its impact 
on the $e$-cloud effect and surface needs to be studied. The issue of synchrotron radiation removal is key to the machine feasibility. 
The first option would be to operate the beam screen in the temperature window of 85-100 K, suitable for vacuum stability. 
However, assuming an efficiency between cryo power and plug power of 10\%, the 2$\times$2~MW power removed at 90~K 
implies a requirement of approximately 2$\times$20~MW for the cryogenic plant. Since the cryogenic power for the full ring 
is of the order of 120~MW (for 1.9~K magnets) or 40~MW (for 4.2~K magnets), the additional power consumption of 40~MW for removal 
of synchrotron radiation remains acceptable. However, the problem of keeping the beam screen well insulated  (at 100~K) 
from the vacuum pipe (at 1.9 or 4.2~K) while removing longitudinally the 37~W/m inside the vacuum pipe represents a major 
challenge. The cooling pipe of the beam screen may become too large and require an increased magnet aperture, beyond the 
40~mm, that has been the guideline for the HE and VHE-LHC magnet concepts, so far.
Another possible solution is to investigate the possibility of letting the radiation escaping from the beam pipe and 
intercept it with cold fingers (or photons stops) at a temperature to be optimized in the range between 80~K and room temperature, 
as proposed  in the US VLHC study~\cite{Pivi:2001uk}. In the case the cold finger is kept at 80~K, the cryogenic load of 40~MW will 
remain, however the issue of local removal of the heat along the beam screen is avoided. The magnet aperture might still need to be 
increased. 
A detailed study and optimization, which includes the length of the main dipoles, should establish the best solution. Although the 
problem of synchrotron radiation is challenging, it should not been considered to be a showstopper. The benefit of the strong 
damping due the synchrotron radiation at 50~TeV/beam should be underlined. Transverse damping time becomes about 30 minutes, 
to be compared to one day at the LHC, and this will greatly contribute to solve the issue of beam stability due to impedance 
or other collective effects.  The coating of the beam screen with HTS, 
as proposed for a HE-LHC, may be not required for a long tunnel. The cryogenic system will have to be much more powerful than that 
installed at the LHC, because of the tunnel length and of the dumped SR power. It will represent a cost driver and will seriously impact 
the facility power consumption. A very preliminary estimate of the electrical power needed for the cryo-system is in the
range of 150-200~MW.

At this time, it is important to ensure a significant US participation in the study of a high energy proton collider in a large tunnel. 
The study will capitalize on experience in the US community, will inform directions for an expanded US technology 
reach and guide the long-term roadmap with strong synergies with the muon and high intensity machines.

\section*{Concluding recommendations}

The study group agrees on the following recommendations for the future development of hadron collider research and technology:

\begin{enumerate}
\item  The full exploitation of the LHC is the highest priority of the energy frontier, hadron collider program.
 A vigorous continuation of the LARP program leading to U.S. participation in the HL-LHC construction project 
is crucial to the full realization of the LHC potential to deliver discovery physics to the thousands of U.S. 
researchers engaged in its experiments;

\item Focused engineering development is no substitute for innovative R\&D. A systematic long-range research in magnet 
materials and structures needs to be vigorously continued. The crucial hadron collider R\&D areas have all substantial implications 
for lepton colliders and the intensity frontier. In particular, the high-field magnet development is key to the muon accelerator 
program and will extend the US R\&D portfolio.

\item  High energy hadron colliders have produced, and will produce over the next decade or more, some of the most fruitful 
collider physics research. In order to maximize the potential for US HEP, this collider program requires a long-term in-depth 
relation of US laboratories and universities with CERN and other international partners. Following up on the previous VLHC 
studies and in view of the renewed interest in a $\sim$100~TeV proton collider, the participation of the US as a leading 
contributor in the design study of a high energy hadron collider in a large tunnel, being initiated by CERN, is strongly 
encouraged and will significantly benefit the definition of the long-term roadmap for collider particle physics.
\end{enumerate}

\section*{Acknowledgments}

We greatly appreciated the contributions of the colleagues who participated to the activities of the Frontier Capabilities working 
group. We especially thank those who contributed white papers, D.~d'Enterria and E.~Todesco for providing some of the figures, 
G.~Apolllinari, P.C.~Bhat, G.L.~Sabbi and F.~Zimmermann for reviewing this report. The work was partially supported by the DOE OHEP.

\end{document}